\PassOptionsToPackage{numbers,sort&compress}{natbib}
\documentclass[final, 3p, times]{elsarticle}

\usepackage{amsmath}
\usepackage{amssymb}
\usepackage{hyperref}
\usepackage{url}

\hypersetup{colorlinks=true}
\makeatletter
\def\ps@pprintTitle{%
  \let\@oddhead\@empty
  \let\@evenhead\@empty
  \def\@oddfoot{\centerline{\thepage}}%
  \let\@evenfoot\@oddfoot}
\makeatother

\makeatletter
\newcommand{\appendixlabel}[1]{%
  \begingroup
  \protected@edef\@currentlabel{\@Alph\c@section}%
  \label{#1}%
  \endgroup
}
\makeatother

\newcommand{\singlecolumnwidth}{0.475\textwidth}
\newcommand{\singlepluscolumnwidth}{0.65\textwidth}
\newcommand{\doublecolumnwidth}{\textwidth}
\newcommand{\tableformat}{\centering\small}
\newcommand{\routeclosure}[1]{\par\medskip\noindent\textbf{#1.}\quad}

\begin{document}

\title{VEQ: a fast parametric Grad--Shafranov solver for fixed-boundary tokamak equilibria with flexible source profiles}


\author[1]{Ruohan Zhang}

\author[2,3]{Huasheng Xie\corref{cor1}}
\ead{huashengxie@gmail.com}

\author[4]{Yueyan Li}

\author[5]{Weiqi Meng}

\author[1]{Feng Wang}

\author[1]{Zhengxiong Wang\corref{cor1}}
\ead{zxwang@dlut.edu.cn}

\affiliation[1]{organization={Key Laboratory of Materials Modification by Beams of the Ministry of Education, School of Physics, Dalian University of Technology},
    city={Dalian},
    postcode={116024},
    country={China}}

\affiliation[2]{organization={Beijing VeloAlpha Technology Co., Ltd.},
    city={Beijing},
    postcode={100080},
    country={China}}

\affiliation[3]{organization={ENN Science and Technology Development Co., Ltd.},
    city={Langfang},
    postcode={065001},
    country={China}}

\affiliation[4]{organization={School of Science, Tianjin University of Technology and Education},
    city={Tianjin},
    postcode={300222},
    country={China}}

\affiliation[5]{organization={School of Physics, Nankai University},
    city={Tianjin},
    postcode={300072},
    country={China}}

\cortext[cor1]{Corresponding authors.}

\begin{abstract}
    Veloce EQuilibrium (VEQ) is a compact parametric framework for tokamak modeling workflows that repeatedly query continuous fixed-boundary equilibria at low latency. The \texttt{VEQPy} implementation evaluated here is an axisymmetric fixed-boundary Grad--Shafranov solver whose main solve enforces a variationally induced projected residual. Its active unknowns are MXH-type flux-surface harmonics and shifted-Chebyshev coefficients for radial profile and source closures. Six input routes accept pressure-gradient, toroidal-field-function, poloidal-flux-gradient, enclosed toroidal current, current-density and safety-factor information through route-specific closures, while all routes map to the same finite-dimensional residual operator. Controlled tests show route consistency for smooth, mutually compatible inputs generated from a common reference equilibrium. For Pareto-selected reduced configurations in three G-EQDSK cases, the most accurate selected rows correspond to a D-shaped case (9 active parameters, minor-radius-normalized shape error $1.4\times 10^{-3}$, solve-only median 1.6 ms), an H-mode case (65, $1.1\times 10^{-3}$, 19 ms), and an X-point case treated as a smoothed fixed-boundary representation of a diverted boundary (94, $1.9\times 10^{-3}$, 15 ms). Sampled pointwise strong-form Grad--Shafranov diagnostics show that enriching the active representation mainly improves interior force balance, whereas the global RMS and maximum values for the H-mode and X-point cases remain dominated by near-boundary contributions. In an isolated one-dimensional transport-geometry coupling test against the target geometry read from G-EQDSK, the temperature-profile response remains below about one percent. These results support using VEQ for repeated equilibrium-geometry queries, provided that pointwise diagnostics are retained to screen cases requiring boundary refinement, local correction or higher-fidelity equilibrium solves.
\end{abstract}

\begin{keyword}
    tokamak equilibrium \sep Grad--Shafranov equation \sep fixed-boundary equilibrium \sep projected residual formulation \sep flux-surface parameterization
\end{keyword}

\maketitle

\section{Introduction}
\label{sec:introduction}

Magnetohydrodynamic (MHD) equilibrium calculations provide the magnetic geometry and force-balance state used by transport, stability, heating, current-drive, control and integrated-modeling workflows. In many such workflows, the solved state is not the final analysis product but an intermediate object queried repeatedly for flux-surface geometry, metric coefficients, current and safety-factor diagnostics, profile-coordinate transformations and transport-geometry factors. This role is typical of integrated-modeling environments such as IMAS, RAPTOR and SuperCode, where equilibrium information is passed to downstream modules \cite{Imbeaux2015,Felici2018,Haney1992}. It imposes requirements that differ from those of a one-time high-fidelity reconstruction: the representation must retain enough two-dimensional shaping and force-balance information to be useful beyond analytic closures, while remaining compact enough for parameter scans, profile preprocessing and control-oriented iterations. Analytic and engineering closures are inexpensive but usually provide limited residual diagnostics; full equilibrium or reconstruction workflows provide richer solver-native outputs but can be too costly or infrastructure-heavy for inner-loop use. These requirements motivate the target of this paper: a continuous fixed-boundary representation that preserves useful two-dimensional shaping and residual diagnostics while remaining inexpensive to regenerate.

For axisymmetric toroidal plasmas, the equilibrium problem is commonly expressed through the Grad--Shafranov equation \cite{Shafranov1966,GradRubin1958,Kruskal1958,Greene1971}. Established reconstruction tools and fixed-boundary Grad--Shafranov solvers provide high-fidelity numerical equilibria \cite{Lao1985,Lutjens1996,Palha2016,SanchezVizuet2019} and are the appropriate choice when the main requirement is a solver-native force-balance residual or a reconstruction workflow tied to experimental constraints. Fast high-order fixed-boundary solvers, including smooth-boundary approaches such as those of Pataki et al.\ and ECOM \cite{Pataki2013,Lee2015}, address closely related accuracy objectives within their own boundary and solution representations. Pseudo-spectral collocation solvers, such as DESC in the three-dimensional setting \cite{Dudt2020}, evaluate nonlinear force-balance equations on collocation grids for variables represented in a solver-native spectral basis. At the opposite end, low-order engineering closures and systems tools such as METIS \cite{Artaud2018}, together with Miller-type shape models \cite{Miller1998,Arbon2021,Snoep2023}, support rapid scenario modeling by deliberately compressing the geometry. The gap considered here lies between these regimes: once the plasma boundary and profile information are specified, solve the fixed-boundary equilibrium in a compact continuous representation that is geometrically richer than analytic closures, while remaining lighter and more reusable than a full solver-native equilibrium or reconstruction pipeline.

VEQ, an abbreviation of ``Veloce EQuilibrium'', is the compact parametric framework considered in this work and is designed for this intermediate regime. The implementation benchmarked here, released as \texttt{VEQPy}, is the fixed-boundary, axisymmetric realization used for the numerical tests below; the broader VEQ framework is not limited to this benchmark scope. A related preliminary VEQ-R study applies a similar compact spectral philosophy to rotating spherical-torus equilibria \cite{Li2026VEQR}. In the realization studied here, the active unknowns are coefficients of MXH-type flux-surface harmonics and source-related radial profiles, rather than nodal values of a grid-based poloidal-flux map. These coefficients define a finite-dimensional representation of continuous nested-surface equilibria. The main solve selects a state in this manifold by enforcing a variationally induced projected Grad--Shafranov residual. Sampled pointwise strong-form residuals, including the point-collocation least-squares polish discussed in Appendix~\ref{app:collocation-comparison}, diagnose or refine the same state; they do not redefine the primary variational solve. VEQ is therefore complementary to high-fidelity equilibrium and reconstruction solvers: the present realization is intended as a repeated-query equilibrium-geometry provider with an explicit projected residual and sampled pointwise strong-form diagnostics, not as a universal replacement for workflows whose primary requirement is pointwise residual certification in solver-native variables.

The formulation builds on our recent compact MXH--Chebyshev representation of MHD equilibria \cite{Xie2026arxiv}. That work asked how compactly a known equilibrium could be fitted; here the same ansatz becomes the finite-dimensional unknown of a nonlinear parametric Grad--Shafranov solve. The active unknowns include shape coefficients and, depending on the input route, profile or source-related coefficients. The shape residuals are derived from the variational form of the Grad--Shafranov equation using test functions induced by constrained variations of the flux-surface ansatz. Thus the shape block is a variationally structured projection rather than an independently prescribed weighted-residual closure. This finite-dimensional viewpoint is related to earlier moment and direct-variational equilibrium methods \cite{Lao1981,Lao1982,Lao1984,Haney1995,Ludwig1995}, but the unknown here is a continuous MXH--Chebyshev flux-surface representation in the open-source implementation evaluated below.

A second practical requirement is compatibility with the different one-dimensional quantities exposed by analytic studies, integrated-modeling tools, reconstruction files and control-oriented models. These inputs may be pressure gradients, toroidal-field-function derivatives, poloidal-flux gradients, enclosed toroidal current, current-density profiles or safety-factor profiles. In the present implementation, the alternatives are represented by six profile-input routes, denoted PF, PP, PI, PJ1, PJ2 and PQ. The selected route maps the supplied quantities to a common normalized source representation used by the same projected residual operator. Route flexibility therefore changes which quantities are supplied and which one-dimensional closure relations are used, not the underlying equilibrium equation. This separation between input semantics and residual assembly makes it possible to compare mutually compatible routes within one finite-dimensional residual system, and to distinguish input-route consistency from geometric approximation error and residual localization.

This paper makes four contributions. First, we define a finite-dimensional parametric MXH--Chebyshev flux-surface representation and derive the variationally induced projected Grad--Shafranov residual used by the main solve, including the convective variation associated with flux-surface motion. Second, we introduce route-specific source closures that map heterogeneous pressure-gradient, poloidal-flux-gradient, current-based and safety-factor inputs to a common finite-dimensional residual system. Third, we report both the projected residual norm minimized by the nonlinear solve and sampled pointwise strong-form Grad--Shafranov residual diagnostics, thereby distinguishing convergence of the reduced system from pointwise force-balance screening. Fourth, we benchmark the present fixed-boundary implementation on Solov'ev-, CHEASE- and EFIT-derived G-EQDSK cases, quantifying input-route consistency, flux-surface reconstruction error, reduced-order geometry--cost trade-offs, residual localization and repeated-solve latency.

Accordingly, the numerical tests are organized around the intended VEQ use case rather than around a single cross-code speed or residual ranking. A controlled reference equilibrium is first used to test consistency among profile-input routes generated from mutually compatible smooth profiles. Three representative fixed-boundary G-EQDSK cases are then considered: a D-shaped file generated from an ITER-like Solov'ev analytic solution, a CHEASE H-mode export and an EFIT X-point export. For these cases we report direct root-mean-square flux-surface reconstruction errors, projected residual norms, sampled standard-form residual statistics and reduced-order Pareto trends. The pointwise residual maps distinguish geometry-dominated cases from cases in which residual statistics remain concentrated near the boundary. An isolated one-dimensional transport-geometry coupling test further examines how errors in VEQ transport-geometry factors propagate through a representative flux-surface-averaged transport operator. Reported timings are post-setup solve-only timings; setup, workspace allocation, just-in-time compilation, file processing and file-to-representation conversion are excluded and treated separately in the benchmark protocol.

\section{Parametric equilibrium representation and variational formulation}
\label{sec:formulation}

This section builds the finite-dimensional equilibrium representation used by the main solve. We first fix the flux-surface coordinate and normalization conventions, then introduce the MXH--Chebyshev surface parameterization and the metric factors evaluated from it. These ingredients lead to the transformed Grad--Shafranov residual and, finally, to the projected finite-dimensional residual system enforced by the solver.

\subsection{Flux-surface coordinates and conventions}
\label{sec:flux-coords}

The implementation benchmarked here treats axisymmetric fixed-boundary tokamak equilibria in nested flux-surface coordinates $(\rho,\theta,\phi)$. The radial label $\rho\in[0,1]$ labels nested magnetic surfaces, with the last closed flux surface (LCFS) at $\rho=1$ and $\rho=0$ corresponding to the magnetic-axis limit. In this implementation, $\rho$ is the normalized radial label of the parametric surface map. It is not assumed to coincide with a normalized poloidal-flux coordinate exported by an equilibrium file; instead, the normalized poloidal flux $\hat{\psi}$ is reconstructed as a monotone function of $\rho$. The coordinate convention follows the COCOS-1 orientation used by the implementation \cite{Sauter2013}: $\theta$ is measured clockwise in the poloidal plane, and the toroidal coordinate follows the corresponding sign convention in the geometric map.

The equilibrium state is represented by a parametric flux-surface map
\begin{equation}
    (\rho,\theta) \mapsto \left[R(\rho,\theta), Z(\rho,\theta)\right],
\end{equation}
together with one-dimensional profiles defined on the same radial label. The physical poloidal-flux derivative is decomposed into a flux scale and a normalized radial profile,
\begin{equation}
    \psi_\rho = \alpha_2 \hat{\psi}_\rho,
\end{equation}
where $\alpha_2$ sets the physical flux scale. The normalized poloidal flux $\hat{\psi}$ is obtained by radial integration when needed. Source amplitudes are separated in the same way through a source scale $\alpha_1$. This scale separation allows different input routes to prescribe different combinations of source profiles and integral constraints while still being converted to a common normalized representation for residual assembly.

In the fixed-boundary realization studied here, the LCFS shape is prescribed. The unknowns are therefore not nodal values of a grid-based $\psi(R,Z)$ field, but active coefficients of radial profiles that deform the interior family of flux surfaces, together with route-dependent flux- or current-based profile coefficients when required. Later comparisons on prescribed normalized poloidal-flux surfaces use the monotone map $\hat{\psi}(\rho)$ to locate the corresponding radial labels. This convention is the basis for the compact nonlinear solve used in the rest of the paper.

\subsection{MXH--Chebyshev parametric representation}

The flux-surface geometry is parameterized through an MXH-type distorted poloidal angle, following the use of low-order Miller-type shaping and its higher-order extensions for compact flux-surface representation and constrained flux-surface parameterization \cite{Miller1998,Arbon2021,Snoep2023}. We define
\begin{equation}
    \bar{\theta}(\rho,\theta)
    = \theta + c_0(\rho)
    + \sum_{m=1}^{M} c_m(\rho)\cos(m\theta)
    + \sum_{n=1}^{N} s_n(\rho)\sin(n\theta),
\end{equation}
and represent the flux surfaces by
\begin{align}
    R(\rho,\theta) & = R_0 + a\left[h(\rho) + \rho\cos\bar{\theta}(\rho,\theta)\right], \\
    Z(\rho,\theta) & = Z_0 + a\left[v(\rho) - \rho\kappa(\rho)\sin\theta\right].
\end{align}
Here $R_0$ and $Z_0$ denote the reference center of the prescribed LCFS used in the boundary parameterization, and $a$ sets the corresponding minor-radius normalization. These constants parameterize the plasma cross-section, not the machine geometry. Magnetic-axis displacement and interior shaping are represented by the radial profile functions. The functions $h$, $v$, $\kappa$, $c_0$, $c_m$ and $s_n$ describe radial shift, vertical shift, elongation, angular offset and higher-order shaping. Low truncation orders recover familiar interpretable shape controls, while higher orders increase representational flexibility without changing the residual-operator structure.

The radial dependence of the shape profiles is expanded in shifted Chebyshev polynomials \cite{Trefethen2000}. With
\begin{equation}
    \xi = 2\rho^2 - 1,
\end{equation}
the basis satisfies
\begin{equation}
    T_0(\xi)=1,\quad T_1(\xi)=\xi,\quad T_{l+1}(\xi)=2\xi T_l(\xi)-T_{l-1}(\xi).
\end{equation}
The radial profile families are written as
\begin{align}
    h(\rho)      & = (1-\rho^2)\sum_{l=0}^{L_h} h_l T_l(\xi),                                      \\
    v(\rho)      & = (1-\rho^2)\sum_{l=0}^{L_v} v_l T_l(\xi),                                      \\
    \kappa(\rho) & = \kappa_a + (1-\rho^2)\sum_{l=0}^{L_\kappa} \kappa_l T_l(\xi),                 \\
    c_0(\rho)    & = c_{0a} + (1-\rho^2)\sum_{l=0}^{L_{c0}} c_{0l}T_l(\xi),                        \\
    c_m(\rho)    & = \rho^{K_m}\left[c_{ma} + (1-\rho^2)\sum_{l=0}^{L_{cm}} c_{ml}T_l(\xi)\right], \\
    s_n(\rho)    & = \rho^{K_n}\left[s_{na} + (1-\rho^2)\sum_{l=0}^{L_{sn}} s_{nl}T_l(\xi)\right],
\end{align}
where the $c_m$ and $s_n$ expressions apply for $m\geq1$ and $n\geq1$, respectively. The coefficients carrying the subscript $a$ prescribe the imposed boundary values at $\rho=1$. The factors $(1-\rho^2)$ make the optimized interior corrections vanish at the fixed boundary.

The powers $\rho^{K_m}$ and $\rho^{K_n}$ regularize higher Fourier harmonics near the magnetic axis, consistent with regularity constraints on Fourier representations in cylindrical coordinates \cite{Lewis1990}. The capped convention $K_m=\min(m,2)$ and $K_n=\min(n,2)$ used in the preceding arXiv paper \cite{Xie2026arxiv} remains a supported modeling choice but is not used below. Unless otherwise stated, the calculations use the regularity-motivated default $K_m=m$ and $K_n=n$, for which higher poloidal harmonics decay rapidly toward the magnetic axis. The exponent choice is independent of both the maximum representable Fourier order and the effective active order used in a particular solve.

\begin{figure}[tb]
    \centering
    \includegraphics[width=\doublecolumnwidth]{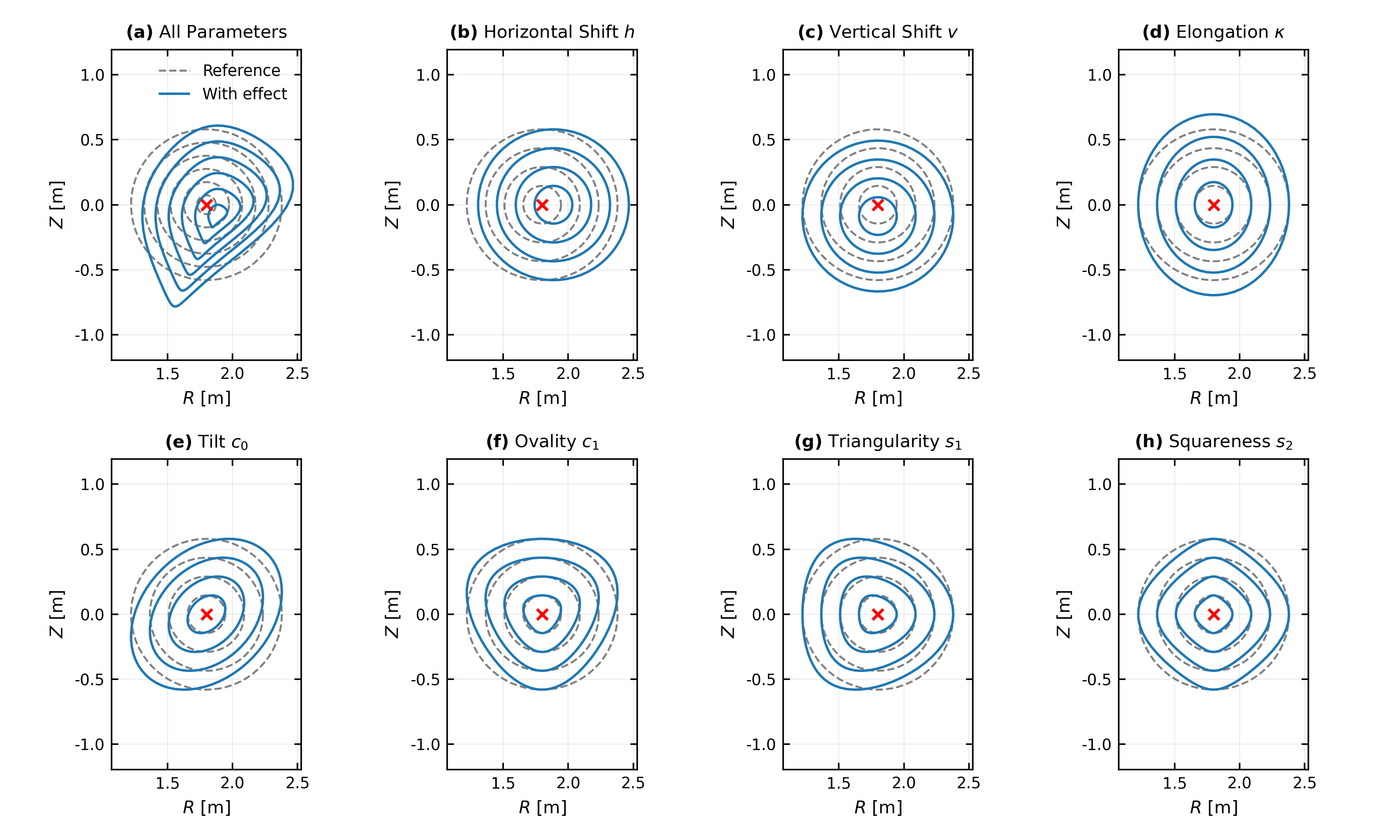}
    \caption{Geometry and profile-basis functions illustrating deformations of the reference flux surfaces.}
    \label{fig:shaping-coefficients}
\end{figure}

Figure~\ref{fig:shaping-coefficients} illustrates the main shaping coefficients and representative radial basis functions. Changing the active order changes the dimension of the nonlinear system, but not the form of the residual operator derived below.

\subsection{Geometry, metrics and admissibility}

The residual assembly and route-dependent source reconstruction use metric quantities computed from the parametric map $(\rho,\theta)\mapsto(R,Z)$. In the poloidal cross-section, the covariant metric is
\begin{equation}
    \left[
        \begin{array}{cc}
            g_{\rho\rho}   & g_{\rho\theta}   \\
            g_{\theta\rho} & g_{\theta\theta}
        \end{array}
        \right]
    =
    \left[
        \begin{array}{cc}
            R_\rho^2+Z_\rho^2               & R_\rho R_\theta+Z_\rho Z_\theta \\
            R_\rho R_\theta+Z_\rho Z_\theta & R_\theta^2+Z_\theta^2
        \end{array}
        \right],
\end{equation}
and the two-dimensional Jacobian is
\begin{equation}
    J = R_\theta Z_\rho - R_\rho Z_\theta.
\end{equation}
After integration over the ignorable toroidal angle, the volume element used below is
\begin{equation}
    \mathrm{d}V = 2\pi JR\,\mathrm{d}\rho\,\mathrm{d}\theta.
\end{equation}

Several one-dimensional geometric factors are obtained by integrating over the poloidal angle. The cross-sectional area derivative and toroidal volume derivative are
\begin{align}
    S_\rho & = \int_0^{2\pi} J\,\mathrm{d}\theta,      \\
    V_\rho & = 2\pi\int_0^{2\pi} JR\,\mathrm{d}\theta.
\end{align}
The geometry factors used in the current and safety-factor relations are written in normalized form as
\begin{align}
    \hat{K}      & = \frac{1}{2\pi}\int_0^{2\pi}\frac{g_{\theta\theta}}{JR}\,\mathrm{d}\theta, \\
    \hat{L}_\rho & = \frac{1}{2\pi}\int_0^{2\pi}\frac{J}{R}\,\mathrm{d}\theta.
\end{align}
These quantities are evaluated from the parametric geometry at each residual state and are reused by the source closures and residual equations introduced below.

The admissible parameter domain is restricted by the regularity of the nested-surface map. In particular, the normalized poloidal-flux label must remain monotone and the coordinate map must not self-intersect:
\begin{equation}
    \hat{\psi}_\rho >0, \quad J>0
\end{equation}
away from the magnetic axis. In the numerical implementation, these conditions are monitored during route construction, initialization and residual evaluation. Iterates that lose flux monotonicity or produce invalid metric factors are rejected or treated as failed solves rather than accepted as equilibria. Near the magnetic axis, $J$ vanishes in the usual polar-coordinate sense. Singular metric divisions are therefore evaluated through regular limiting behavior or avoided at the degenerate axis point in diagnostic grids.

\subsection{Grad--Shafranov residual from the variational equation}

We next derive the local transformed Grad--Shafranov residual that will be projected onto the finite-dimensional representation in the following subsection. The derivation starts from the equilibrium variational equation rather than from an independently chosen weighted residual. This places the formulation in the lineage of variational and moment approaches to Grad--Shafranov equilibria \cite{Lao1981,Lao1984,Ludwig1995}. Consider the axisymmetric energy functional
\begin{equation}
    \mathcal{L}=\int L\,\mathrm{d}V,
    \quad
    \mathrm{d}V=2\pi JR\,\mathrm{d}\rho\,\mathrm{d}\theta,
\end{equation}
with Lagrangian density
\begin{equation}
    L = \frac{|\nabla\psi|^2}{2\mu_0R^2}
    - \left(\frac{F^2}{2\mu_0R^2}+P\right).
\end{equation}
For a flux variation at fixed coordinates, integration by parts gives
\begin{equation}
    \delta\mathcal{L}
    =\,2\pi\int_0^1\!\int_0^{2\pi}
    \left[
        JR\frac{\partial L}{\partial\psi}
        -\frac{\partial}{\partial\rho}\left(JR\frac{\partial L}{\partial\psi_\rho}\right)
        -\frac{\partial}{\partial\theta}\left(JR\frac{\partial L}{\partial\psi_\theta}\right)
        \right]\delta\psi\,
    \mathrm{d}\theta\,\mathrm{d}\rho,
\end{equation}
where boundary terms vanish for admissible fixed-boundary variations. We define the transformed Grad--Shafranov residual density by
\begin{equation}
    \mathcal{G}
    =-\mu_0\left[
        JR\frac{\partial L}{\partial\psi}
        -\frac{\partial}{\partial\rho}\left(JR\frac{\partial L}{\partial\psi_\rho}\right)
        -\frac{\partial}{\partial\theta}\left(JR\frac{\partial L}{\partial\psi_\theta}\right)
        \right].
\end{equation}
Stationarity for arbitrary admissible $\delta\psi$ gives $\mathcal{G}=0$, which is equivalent to the strong Grad--Shafranov equation in the present coordinates. The corresponding standard cylindrical strong-form residual is
\begin{equation}
    \mathcal{G}_{\mathrm{std}}
    =\frac{R}{J}\mathcal{G}
    =\Delta^{*}\psi+FF_\psi+\mu_0R^2P_\psi,
    \quad
    \Delta^{*}\psi
    =R\frac{\partial}{\partial R}\left(\frac{1}{R}\frac{\partial\psi}{\partial R}\right)
    +\frac{\partial^2\psi}{\partial Z^2}.
    \label{eq:standard-gs-residual}
\end{equation}
At sampled non-axis points, $R/J$ is nonsingular for an admissible nested-surface map. Thus $\mathcal{G}=0$ and $\mathcal{G}_{\mathrm{std}}=0$ define the same continuous equilibrium equation away from the coordinate-degenerate magnetic axis. The axis is handled through regular limiting behavior rather than by evaluating $R/J$ at $\rho=0$.

The factor $J/R$ in the transformed residual follows from the Euler--Lagrange operator in flux-surface coordinates: the volume Jacobian $JR$, the $R^{-2}$ magnetic and toroidal-field terms, and the metric expression for $|\nabla\psi|^2$. In particular,
\begin{equation}
    |\nabla\psi|^2
    =\frac{g_{\theta\theta}\psi_\rho^2-2g_{\rho\theta}\psi_\rho\psi_\theta+g_{\rho\rho}\psi_\theta^2}{J^2}.
\end{equation}
Using the flux-surface representation $\psi_\theta=0$, the residual becomes
\begin{equation}
    \mathcal{G}
    = \frac{J}{R}\left(FF_\psi+\mu_0R^2P_\psi\right)
    + \left(\frac{g_{\theta\theta}\psi_\rho}{JR}\right)_\rho
    - \left(\frac{g_{\rho\theta}\psi_\rho}{JR}\right)_\theta .
\end{equation}
The $J/R$ multiplying the source term is therefore part of the transformed Grad--Shafranov operator, not an extra quadrature factor or empirical weighting choice. Here $FF_\psi\equiv F\,\partial F/\partial\psi$ denotes the single toroidal-field source profile in standard Grad--Shafranov notation. The projected residual system uses this transformed coordinate residual density $\mathcal{G}$. For sampled pointwise diagnostics in Section~\ref{sec:residual-distribution}, we report the corresponding standard-form residual $\mathcal{G}_{\mathrm{std}}=(R/J)\mathcal{G}$, the same strong-form density used by the optional point-collocation least-squares polish discussed in Appendix~\ref{app:collocation-comparison}, up to the normalizations introduced by the selected route.

For the normalized source representation, the same residual is decomposed into a source part and a poloidal-flux-gradient part,
\begin{equation}
    \mathcal{G}=\alpha_1\hat{G}_1+\alpha_2\hat{G}_2,
\end{equation}
with
\begin{align}
    \hat{G}_1 & = \frac{J}{R}\left(\hat{FF}_\psi + R^2\hat{P}_\psi\right), \\
    \hat{G}_2 & = \frac{g_{\theta\theta}}{JR}\hat{\psi}_{\rho\rho}
    + \left[\left(\frac{g_{\theta\theta}}{JR}\right)_\rho
        - \left(\frac{g_{\rho\theta}}{JR}\right)_\theta\right]\hat{\psi}_\rho .
\end{align}
The input route, defined in Section~\ref{sec:routes}, determines how $\alpha_1$, $\alpha_2$, $\hat{FF}_\psi$, $\hat{P}_\psi$ and $\hat{\psi}_\rho$ are obtained. After this conversion, all routes use the same local transformed residual form above.

\subsection{Variational projected residuals and profile closures}

Having defined the local transformed residual, we now restrict the admissible variations to those generated by the active coefficients of the parametric representation. For a shape parameter $p_k$, the labeled flux surface itself moves. Since $\psi$ is constant on a labeled surface, the Eulerian flux variation and the surface displacement satisfy the convective-variation identity
\begin{equation}
    \delta_{\mathbf{r}}\psi+\nabla\psi\cdot\delta\mathbf{r}=0,
    \quad
    \delta\mathbf{r}=\sum_k \frac{\partial\mathbf{r}}{\partial p_k}\delta p_k .
\end{equation}
The active shape variations are interior variations: the boundary coefficients are held fixed, and the factors $(1-\rho^2)$ in the optimized corrections make the variations vanish at $\rho=1$. Thus the convective variation moves the interior family of labeled flux surfaces while preserving the prescribed LCFS.

A tangential displacement only reparametrizes a labeled surface and does not add an independent force-balance condition. The projected shape equation is therefore driven by the induced normal flux variation $\nabla\psi\cdot\partial\mathbf{r}/\partial p_k$, with a sign convention that is immaterial for the zero-residual equations when used consistently. Equivalently,
\begin{equation}
    \frac{\partial\psi}{\partial p_k}
    =-\nabla\psi\cdot\frac{\partial\mathbf{r}}{\partial p_k}.
\end{equation}
In the present coordinates, with $\psi=\psi(\rho)$,
\begin{equation}
    \frac{\partial\psi}{\partial R}=-\frac{Z_\theta}{J}\psi_\rho,
    \quad
    \frac{\partial\psi}{\partial Z}=\frac{R_\theta}{J}\psi_\rho.
\end{equation}

Substituting this convective variation into $\delta\mathcal{L}$ gives the shape-parameter stationarity equations
\begin{equation}
    \int_0^1\!\int_0^{2\pi}
    \mathcal{G}\left(
    \frac{\partial\psi}{\partial R}\frac{\partial R}{\partial p_k}
    +
    \frac{\partial\psi}{\partial Z}\frac{\partial Z}{\partial p_k}
    \right)
    \mathrm{d}\theta\,\mathrm{d}\rho=0.
\end{equation}
Define the shape-induced test function
\begin{equation*}
    \chi_k(\rho,\theta)
    =\nabla\psi\cdot\frac{\partial\mathbf{r}}{\partial p_k}
    =\frac{\psi_\rho}{J}
    \left(R_\theta\frac{\partial Z}{\partial p_k}
    -Z_\theta\frac{\partial R}{\partial p_k}\right).
\end{equation*}
Then the shape residual can be written as the quadrature-sampled Petrov--Galerkin condition
\begin{equation}
    \sum_{i=1}^{N_\rho}\sum_{j=1}^{N_\theta}
    w_i^\rho w_j^\theta\,
    \mathcal{G}_{ij}\,\chi_{k,ij}=0,
\end{equation}
where $\mathcal{G}_{ij}=\mathcal{G}(\rho_i,\theta_j)$ and $\chi_{k,ij}=\chi_k(\rho_i,\theta_j)$. The poloidal factor may be taken either as the trapezoidal integration weight $2\pi/N_\theta$ or as the corresponding average $1/N_\theta$; the difference is a common constant in the zero-residual equations. After flattening the residual samples into $\mathbf{g}$ and collecting the shape test functions into $\mathcal{X}_{qk}=\chi_{k,q}$, the shape block has the compact form
\begin{equation}
    \mathcal{X}^{T}\mathcal{W}\mathbf{g}=0.
\end{equation}
This is the projected variational condition used for the shape degrees of freedom. The test functions are not arbitrary monomial or collocation weights; they are the flux variations induced by constrained motion of the parameterized flux surfaces.

The shape projection supplies the residual equations for the active shape coefficients. Some input routes also make $\hat{\psi}$ or $F$ an active one-dimensional unknown. These coefficients are closed by residual moment conditions with basis-induced radial tests. The resulting route-closure equations select the finite-dimensional profile representation entering the common residual. They are not additional geometric variations of the flux surfaces, nor are they obtained by varying the MHD functional with respect to an independent $F$ field. The one-dimensional unknowns use the same shifted-Chebyshev radial basis as the geometry profiles. With the profile form used in the implementation, the normalized poloidal-flux map and an explicit toroidal-field family are parameterized as
\begin{align}
    \hat{\psi}(\rho)
     & = \rho^2\left[1+(1-\rho^2)\sum_{l=0}^{L_{\hat{\psi}}}\hat{\psi}_l T_l(\xi)\right], \\
    F(\rho)
     & = R_0B_0\left[1+(1-\rho^2)^2\sum_{l=0}^{L_F}F_lT_l(\xi)\right]^{1/2},
\end{align}
with $\hat{\psi}_\rho$ obtained by differentiating the first expression. Depending on the route, the toroidal-field profile is represented by the $F$ family, from which the source form needed to infer $FF_\psi$ under the route constraints is constructed. In such cases, the $\hat{\psi}$ and $F$ coefficients are closed by moment projections of the same residual density over the $(\rho,\theta)$ measure, without additional geometric weighting. For example, with basis-induced profile tests,
\begin{align}
     & \int_0^1\!\int_0^{2\pi}
    \mathcal{G}\,\frac{\partial\psi}{\partial\psi_k}
    \,\mathrm{d}\theta\,\mathrm{d}\rho=0, \\
     & \int_0^1\!\int_0^{2\pi}
    \mathcal{G}\,\frac{\partial F}{\partial F_k}
    \,\mathrm{d}\theta\,\mathrm{d}\rho=0 .
\end{align}

After normalization and removal of common constants, the shape-projection equations and the route-dependent profile moment equations form the residual vector assembled by the operator. Let $\mathcal{A}\in\mathbb{R}^{N_\rho N_\theta\times n}$ collect the finite set of shape- and profile-induced test functions used by a given solve. With
\begin{equation*}
    \mathbf{g}(\mathbf{x})
    =
    \bigl[\mathcal{G}(\rho_i,\theta_j;\mathbf{x})\bigr]_{q=1}^{N_\rho N_\theta},
    \qquad q=(i,j),
\end{equation*}
the assembled projected conditions can be written as
\begin{equation}
    \mathcal{A}^{T}\mathcal{W}\mathbf{g}=0.
    \label{eq:finite-projected-residual}
\end{equation}
Equation~\eqref{eq:finite-projected-residual} is a Galerkin orthogonality statement on the finite ansatz-induced test space. It states that the sampled residual lies in the weighted orthogonal complement of the test space; it does not imply $\mathbf{g}=0$. When $N_\rho N_\theta>\mathrm{rank}(\mathcal{A})$, this complement is nontrivial, so nonzero sampled transformed-density residual components are expected unless the exact equilibrium, boundary representation, source closure, differentiation and quadrature are all compatible with the chosen finite-dimensional parameter manifold.

The projected condition should therefore not be interpreted as imposing $\mathcal{G}_{ij}=0$ at every sample point, nor as the normal equation of a sampled point-collocation least-squares objective \cite{Nocedal2006}. The optional point-collocation least-squares polish and the corresponding finite-dimensional comparison are discussed in Appendix~\ref{app:collocation-comparison}.

Thus the solution reported in this paper is obtained from a finite-dimensional projected residual system with a variational shape block and moment-closure profile blocks. It reduces the Grad--Shafranov residual in the directions available to the chosen parametric family while preserving the nested-surface representation, fixed LCFS and axis-regularized harmonic structure encoded in the ansatz. For smooth compatible equilibria, the Fourier/MXH angular representation and shifted-Chebyshev radial profiles provide the usual high-order approximation capacity of spectral expansions \cite{Trefethen2000}; the realized convergence still depends on boundary smoothness, source compatibility and active-order choices. The sampled diagnostics reported below make explicit the residual components outside this finite test space.

\section{Route conventions and common residual assembly}
\label{sec:routes}

Section~\ref{sec:formulation} defined the transformed residual and projected finite-dimensional system. This section specifies the interface between that common system and the heterogeneous one-dimensional inputs used in practice. We first define the supported input routes, normalization conventions and source-coordinate choices that convert supplied profiles into canonical residual variables. We then describe how the route closures, active profile choices and residual blocks form a square map for the nonlinear solver. Finally, we specify what is exported after convergence and how post-solve quantities are reevaluated from the continuous representation.

\subsection{Input routes, normalization constraints and source-coordinate conventions}
\label{sec:input-routes}

The supported routes differ in the one-dimensional quantities supplied as input and in the closures used to recover the canonical residual variables; they do not define different equilibrium models. Analytic studies, integrated-modeling tools, reconstruction files and control-oriented models often provide different subsets of pressure-gradient, toroidal-field, poloidal-flux-gradient, current, current-density or safety-factor information. A route-specific closure converts these heterogeneous inputs to the normalized source representation used by the common residual expression. Thus route flexibility changes the supplied quantities and the one-dimensional closure relation, not the Grad--Shafranov residual equation being assembled.

Table~\ref{tab:input-routes} summarizes the six routes used in this work. PF supplies pressure-gradient and toroidal-field source derivatives. PP supplies pressure-gradient and poloidal-flux-gradient profiles. PI, PJ1 and PJ2 use enclosed toroidal current, flux-surface-averaged toroidal current density and parallel current density, respectively. PQ uses the safety-factor profile.

\begin{table}[tb]
    \caption{Supported input routes and canonical quantities recovered for the common residual. Direct inputs are physical quantities before normalization; recovered quantities are supplied by route closures.}
    \tableformat
    \begin{tabular}{l c c}
        \hline
        Route & Direct input                & Recovered for residual                   \\
        \hline
        PF    & $P_\psi,\;FF_\psi$          & $\psi_\rho$                              \\
        PP    & $P_\psi,\;\psi_\rho$        & $FF_\psi$                                \\
        PI    & $P_\psi,\;I_{\mathrm{tor}}$ & $\psi_\rho,\;FF_\psi$                    \\
        PJ1   & $P_\psi,\;j_{\mathrm{tor}}$ & $I_{\mathrm{tor}},\;\psi_\rho,\;FF_\psi$ \\
        PJ2   & $P_\psi,\;j_{\parallel}$    & $I_{\mathrm{tor}},\;\psi_\rho,\;FF_\psi$ \\
        PQ    & $P_\psi,\;q$                & $\psi_\rho,\;FF_\psi$                    \\
        \hline
    \end{tabular}
    \label{tab:input-routes}
\end{table}

Unhatted symbols in Table~\ref{tab:input-routes} denote physical input quantities before normalization, while hatted symbols denote normalized profiles entering the residual. The word ``recovered'' means that the missing canonical quantity is supplied by the route closure. It does not imply that the quantity is evaluated only once before the nonlinear solve.

Operationally, each route has three layers. First, the user supplies the physical quantities listed in Table~\ref{tab:input-routes}. Second, the finite-dimensional unknown vector contains the shape coefficients and any route-dependent profile coefficients, such as $\hat{\psi}$ or $F$. Third, during each residual evaluation, the route closure reconstructs the missing canonical profiles and scale factors before the common residual expression is evaluated. This separation between supplied data, active unknowns and residual-call reconstruction is why PF($\hat{\psi}$) and PF($\rho$) may differ in source-coordinate evaluation.

In these route definitions, hats denote normalized source or profile quantities entering the common residual, while $\alpha_1$ and $\alpha_2$ restore the physical source and flux scales. The geometric factors $\hat{K}$, $\hat{L}_\rho$, $S_\rho$ and $V_\rho$ are the quantities defined in Section~\ref{sec:formulation}.

The source-coordinate convention determines which source data are fixed during setup and which quantities are reevaluated with the current nonlinear iterate. For $\rho$-coordinate or grid-node inputs, the source samples and interpolation data are fixed during setup, while the geometry-dependent route closure is evaluated during each residual call. For inputs tabulated against the active normalized poloidal flux, such as PF($\hat{\psi}$) and PQ($\hat{\psi}$), the sampled source values also depend on the iterate-dependent map $\hat{\psi}(\rho)$ and are therefore reevaluated during the same residual call.

Because profile data may be supplied as derivatives with respect to either flux or the radial label, the notation below distinguishes flux-derivative and radial-derivative source conventions. Here $FF_\psi$ and $FF_\rho$ denote the flux- and radial-derivative forms of the toroidal-field source $F\,\mathrm{d}F/\mathrm{d}(\cdot)$. The implementation also accepts the equivalent $\rho$-derivative convention used by the one-dimensional route equations. We write the physical scaling as
\begin{align}
    \psi_\rho
     & =\alpha_2\hat{\psi}_\rho,                     \\
    FF_\rho
     & =\alpha_1\alpha_2\hat{FF}_\rho,               \\
    P_\rho
     & =\frac{\alpha_1\alpha_2}{\mu_0}\hat{P}_\rho .
\end{align}
The associated flux-derivative forms are
\begin{align}
    FF_\psi
     & =\alpha_1\hat{FF}_\psi,              \\
    P_\psi
     & =\frac{\alpha_1}{\mu_0}\hat{P}_\psi, \\
    \hat{FF}_\rho
     & =\hat{FF}_\psi\hat{\psi}_\rho,       \\
    \hat{P}_\rho
     & =\hat{P}_\psi\hat{\psi}_\rho .
\end{align}
Thus a route may be supplied in flux-derivative form or in an equivalent radial-derivative form, provided that the route closure supplies the canonical normalized quantities required by the residual. The hat in $\hat{FF}_\psi$ applies to the composite toroidal-field source $FF_\psi$. For a normalized toroidal-field profile $F_n(\hat{\psi})$, $\hat{FF}_\psi=F_n\,\mathrm{d}F_n/\mathrm{d}\hat{\psi}$; it is not a derivative with respect to the physical $\psi$, nor a hat acting on only one factor in $F\,\mathrm{d}F/\mathrm{d}\psi$.

The route conversions reuse the one-dimensional geometric factors defined in Section~\ref{sec:formulation}. The factors $\hat{K}$ and $\hat{L}_\rho$ enter current and safety-factor relations, while $S_\rho$ and $V_\rho$ enter area and volume averages. For any scalar quantity $A$, the volume average used in the beta constraint is
\begin{equation}
    \langle A\rangle_V
    =\frac{\displaystyle\int_0^1\!\int_0^{2\pi}A(\rho,\theta)JR\,\mathrm{d}\theta\,\mathrm{d}\rho}
    {\displaystyle\int_0^1\!\int_0^{2\pi}JR\,\mathrm{d}\theta\,\mathrm{d}\rho},
\end{equation}
and for a flux function this reduces to
\begin{equation}
    \langle A\rangle_V
    =\frac{\displaystyle\int_0^1 A(\rho)V_\rho(\rho)\,\mathrm{d}\rho}
    {\displaystyle\int_0^1 V_\rho(\rho)\,\mathrm{d}\rho} .
\end{equation}

The same geometry factors connect current- and safety-factor-based inputs to the poloidal-flux-gradient profile. The safety factor and enclosed toroidal current satisfy
\begin{equation}
    q=\frac{F\hat{L}_\rho}{\alpha_2\hat{\psi}_\rho},
    \quad
    I_{\mathrm{tor}}=\frac{2\pi\alpha_2}{\mu_0}\hat{K}\hat{\psi}_\rho .
\end{equation}
The corresponding local toroidal current density in canonical PF variables is
\begin{equation}
    j_\phi(\rho,\theta)
    =-\frac{\alpha_1}{\mu_0}
    \left(\frac{\hat{FF}_\psi}{R}+R\hat{P}_\psi\right).
\end{equation}
The flux-surface-averaged toroidal current density and the parallel current density are defined by
\begin{equation}
    j_{\mathrm{tor}}\equiv\langle j_\phi\rangle_S
    =\frac{1}{S_\rho}\frac{\mathrm{d}I_{\mathrm{tor}}}{\mathrm{d}\rho},
    \quad
    \langle A\rangle_S=\frac{1}{S_\rho}\int_0^{2\pi}AJ\,\mathrm{d}\theta,
\end{equation}
where $\langle A\rangle_S$ is the area-weighted surface average. Unless a subscript is shown, angle brackets denote the usual flux-surface average, so that
\begin{equation}
    j_{\parallel}
    =\frac{\langle\mathbf{j}\cdot\mathbf{B}\rangle}{\langle\mathbf{B}\cdot\nabla\phi\rangle}
    =\frac{\alpha_2 F}{\mu_0\hat{L}_\rho}
    \left(\frac{\hat{K}\hat{\psi}_\rho}{F}\right)_\rho .
\end{equation}
These definitions are used to normalize route inputs before residual assembly. Current-based routes infer poloidal-flux-gradient information from the current relations above; the safety-factor route uses the relation for $q$; and active-$F$ routes recover the missing toroidal-field source from the reconstructed $F$ profile. The corresponding one-dimensional closures reduce the supplied inputs to the canonical source set used by the residual expression. Since the main text focuses on route conventions and common assembly, Appendix~\ref{app:route-closures} records the formulas used to compute $\hat{P}_\psi$, $\hat{FF}_\psi$ and $\hat{\psi}_\rho$ for each route. After this conversion, all routes enter the same canonical residual representation.

Integral constraints determine the physical source and flux scales. In the present fixed-boundary implementation, routes may be combined with a total plasma current constraint $I_p$, a toroidal beta constraint $\beta_t$, or both when the supplied inputs provide enough information. The total current constraint is
\begin{equation}
    I_p=I_{\mathrm{tor}}(1)=\int_0^1\!\int_0^{2\pi} j_\phi J\,\mathrm{d}\theta\,\mathrm{d}\rho,
\end{equation}
while the toroidal beta constraint uses
\begin{equation}
    \beta_t = \frac{2\mu_0\langle P\rangle_V}{B_0^2} .
\end{equation}
Here $B_0$ is the reference toroidal magnetic field used in the beta normalization.
These constraints determine the scaling factors $\alpha_1$ and $\alpha_2$ that connect the dimensionless source profiles to physical quantities. This scaling step is deliberately separated from the residual expression, so changing the input convention does not create a different residual equation.

\subsection{Common residual assembly and finite-dimensional representation}

\begin{figure}[tb]
    \centering
    \includegraphics[width=\singlepluscolumnwidth]{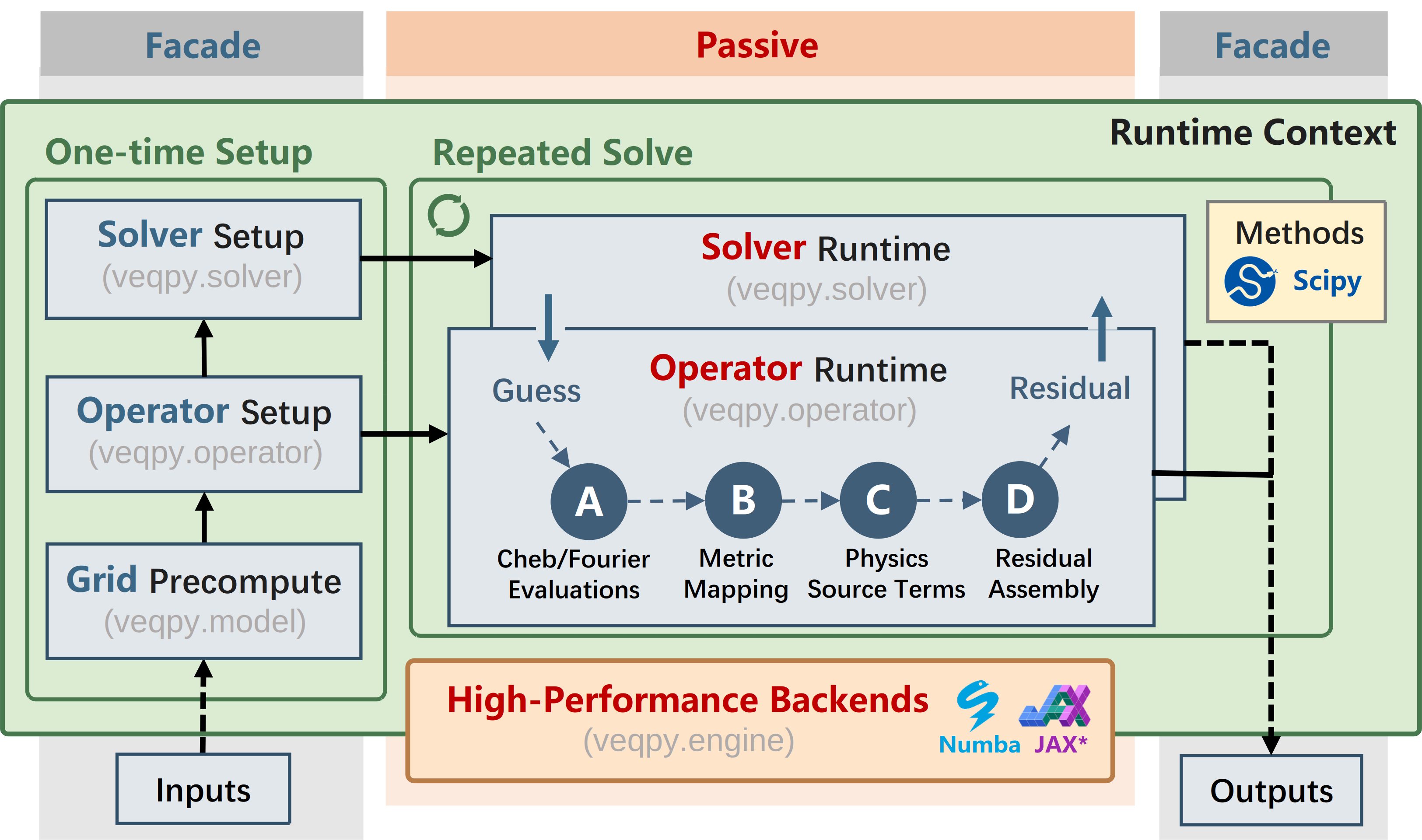}
    \caption{Schematic workflow of the VEQ residual operator and nonlinear solve. One-time setup fixes the case definition, grid-dependent precomputation, operator data and solver configuration. During the post-setup repeated-solve stage, the nonlinear solver supplies a coefficient vector $\mathbf{x}$ and the operator evaluates the common profile--geometry--source--residual pipeline: Stage A updates profile fields, Stage B evaluates geometry and metric factors, Stage C evaluates route-dependent canonical source quantities and scale factors, and Stage D assembles the common residual vector. Precomputed arrays, index maps and compiled kernels are reused across residual calls. After convergence, the solved state is evaluated into output profiles, geometry and diagnostics. The JAX pathway shown schematically is under development, has not been publicly released and is not used in the present timings.}
    \label{fig:operator-pipeline}
\end{figure}

After the route interface has been specified, the implementation converts the route choice, active profile unknowns and selected residual blocks into fixed operator data, as summarized schematically in Figure~\ref{fig:operator-pipeline}. The figure combines the one-time setup path with the post-setup repeated-solve runtime path. The case specification provides the fixed-boundary object, fitted profile families, route and source-coordinate conventions, source arrays and scalar normalization constraints such as $I_p$ or $\beta_t$. Validation, interpolation data, grid-dependent operators and unknown-coefficient lists are fixed during setup. The nonlinear solver therefore sees a uniform square residual map from the finite parameter vector to residual components,
\begin{equation}
    \mathbf{x}=(x_1,\ldots,x_{N})
    \mapsto
    \mathcal{R}(\mathbf{x})=(\mathcal{R}_1,\ldots,\mathcal{R}_{N}).
\end{equation}
For each route, the selected unknown coefficients and residual blocks are matched so that the nonlinear system is square. We reserve the lowercase $r$ for geometric radial distances in the shape-error metric, while $\mathcal{R}$ denotes the residual vector supplied to the nonlinear solver.

Each solve call follows the same profile--geometry--source--residual pipeline: profile fields are updated from $\mathbf{x}$, geometry and metric factors are evaluated, route-dependent canonical source quantities and scale factors are recovered, and the common residual vector is assembled. Separating setup from solve-time evaluation allows nonlinear iterations to reuse precomputed arrays, index maps and compiled routines. In particular, the smooth radial profile representation allows the spectral differentiation and integration matrices used by the route closures to be built once for a given grid and reused during repeated calls.

For a fixed quadrature grid, the two-dimensional geometry and transformed-residual evaluations scale with the number of sampled surface points, while route-closure operations are dominated by one-dimensional profile differentiation, integration and interpolation on the radial grid. The nonlinear system dimension is the number of active coefficients selected for the surface and route-dependent profile families. The reported solve times therefore reflect both the active representation size and repeated use of the common pipeline after setup, precomputation and JIT compilation have been completed.

The nonlinear solve is orchestrated through SciPy nonlinear solvers \cite{Virtanen2020}, with Numba used for the accelerated residual backend \cite{Lam2015}. The evaluator operates on precomputed arrays and index tables; shifted-Chebyshev tables and radial differentiation/integration matrices provide profile values, derivatives and route-specific closures. Radial quadrature uses Gauss weights mapped to $[0,1]$, and the poloidal direction uses equally spaced Fourier/trapezoidal sums. These are implementation choices for the finite-dimensional projected residual. Solver initialization, scaling and acceptance policies wrap the common evaluator without changing the residual definition. Because route-specific source recovery is isolated in Stage C, the residual assembly itself is shared by all routes.

\subsection{Exported equilibrium state and post-solve resampling}

After convergence, the solved object remains the coefficient vector $\mathbf{x}$. For output, the operator evaluates this coefficient vector on a chosen grid and exports profiles, geometry, source quantities and diagnostics. The converged one-dimensional profiles and flux-surface map define the exported continuous form. Thus figures, error metrics and derived quantities ($I_p$, $q$, $s$, $j_{\mathrm{tor}}$, $j_{\parallel}$, source terms) are reevaluated from those profiles and the parametric geometry, rather than obtained by interpolating a stored rectangular $\psi(R,Z)$ array frozen at the solve grid. Diagnostic grids denser than the solve grid use this same continuous reevaluation.

\section{Benchmark protocol and input-route consistency tests}
\label{sec:benchmark}

With the route interface and common residual assembly defined above, this section specifies the benchmark protocol used before the representative G-EQDSK applications. The purpose is to isolate route conversion, resolution dependence and repeated-solve timing under controlled conditions. We first construct a controlled reference equilibrium from smooth prescribed sources and use it to generate mutually compatible route inputs. We then specify initialization policies, error metrics and timing conventions. Finally, we report the resolution dependence and multi-route consistency tests used to verify the route-conversion and residual-assembly workflow.

\subsection{Controlled reference equilibrium and source profiles}

\begin{figure}[tb]
    \centering
    \includegraphics[width=\singlecolumnwidth]{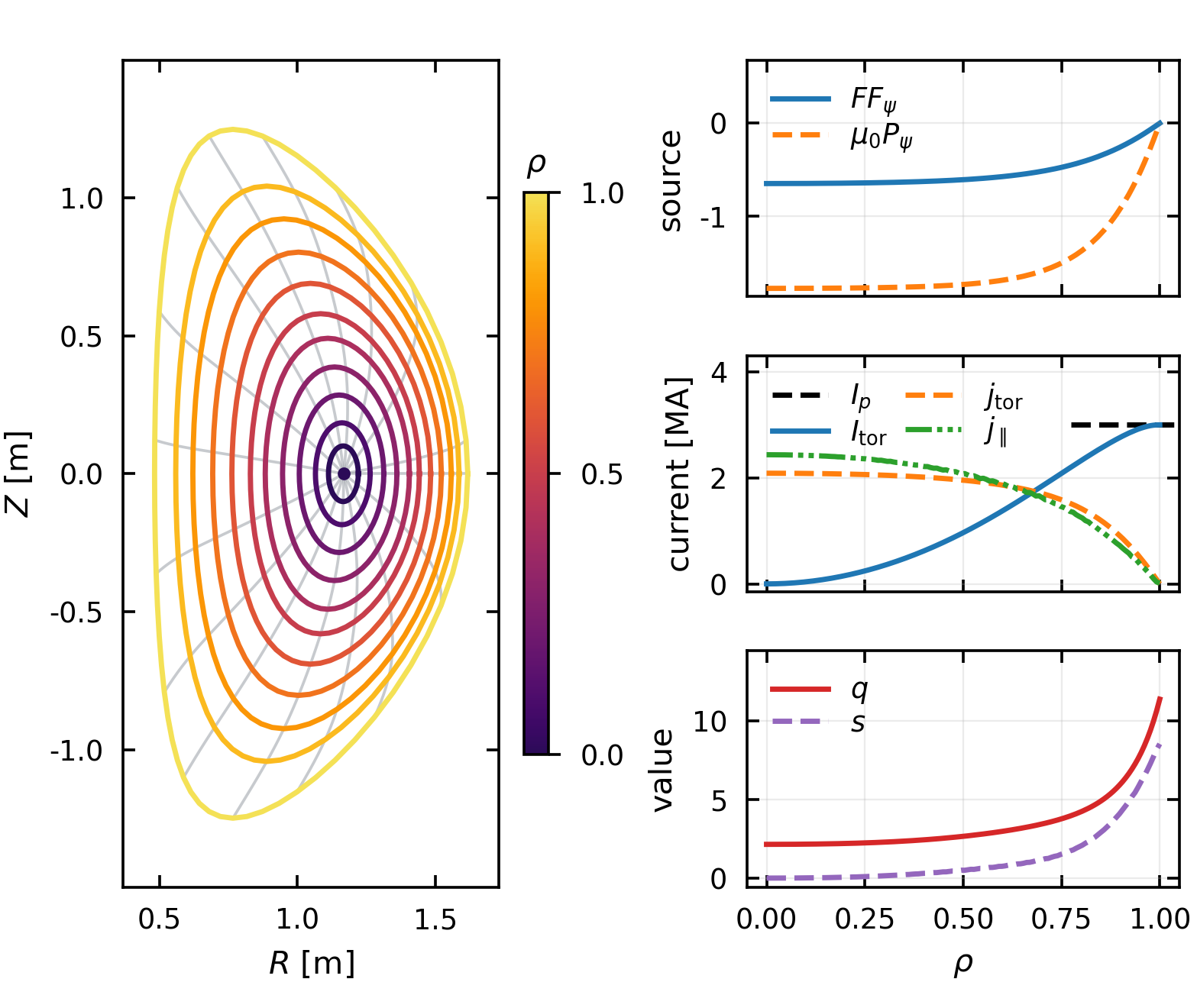}
    \caption{Controlled PF-route reference equilibrium used for the route-consistency benchmark. The post-solve evaluated state provides the reference flux surfaces and derived one-dimensional diagnostics used to generate compatible route inputs.}
    \label{fig:demo-equilibrium}
\end{figure}

The first numerical tests are controlled consistency benchmarks rather than comparisons with an experimental reconstruction. They ask whether mutually compatible physical information supplied through different input routes recovers the same fixed-boundary parametric equilibrium. A high-resolution PF-route reference equilibrium is first constructed from prescribed smooth source profiles and a fixed boundary. The resulting evaluated state, shown in Figure~\ref{fig:demo-equilibrium}, provides the reference geometry and derived one-dimensional profiles used to generate compatible inputs for the other routes. The steep edge safety factor and magnetic shear make this case sensitive to near-boundary interpolation and profile recovery, which is useful for testing the route-closure workflow.

\subsection{Benchmark initialization policies}
\label{sec:initialization-policies}

Initialization can affect nonlinear convergence and timing, but it is not part of the VEQ residual definition. We therefore specify it as part of the benchmark protocol. The controlled reference, resolution and route-consistency benchmarks in Figures~\ref{fig:demo-equilibrium}--\ref{fig:multi-route-consistency} use the all-zero active-coefficient initial state. The representative reconstruction and Pareto benchmarks in Figures~\ref{fig:high-order-reconstructions}--\ref{fig:pareto-analysis} instead use a cold-start homothetic initialization for every timing sample, with warm starts disabled.

In this policy, a low-cost nested-surface seed is formed from the fixed-boundary shape. For nonuniform source profiles, the leading Shafranov-shift coefficient is initialized by the estimate $0.66a/R_0$. Each active cosine or sine family is initialized through only its leading interior coefficient, so that the corresponding profile connects smoothly from regular magnetic-axis behavior to the boundary offset. All remaining active coefficients start at zero. For uniform Solov'ev-like source profiles, this estimate vanishes and the homothetic seed reduces to the zero state. This initialization is only a solver seed; it is not an additional fitted model or a continuation path.

\subsection{Error definitions and timing protocol}
\label{sec:error-timing}

Two error-symbol conventions are used throughout the diagnostics. Scalar quantities are reported with the relative error
\begin{equation}
    \Delta_y = \frac{|y-y^{\mathrm{ref}}|}{|y^{\mathrm{ref}}|},
\end{equation}
for the nonzero scalar diagnostics used here. For the route-consistency diagnostics, $y\in\{I_p,\beta_t,q_{95}\}$, and $q_{95}$ is obtained by interpolating the exported safety-factor profile at $\hat{\psi}=0.95$. Because this point lies close to the plasma edge, $q_{95}$ is more sensitive to radial resolution and interpolation than volume-integrated quantities.

RMS-type coefficient and geometric errors are denoted by $E$. The coefficient metric $E_{\mathrm{coeff}}$ is the RMS difference of the shared active shape-parameter entries $p_i$, with no reference-amplitude normalization.

For two-dimensional shape comparisons, G-EQDSK geometries and VEQ reconstructions are sampled on common normalized poloidal-flux surfaces and common geometric poloidal angles. In the representative cases below, these comparison geometries are Solov'ev, CHEASE and EFIT G-EQDSK files. Let $\chi$ denote the geometric polar angle about the magnetic axis, distinct from the parametric poloidal coordinate $\theta$ used in the MXH representation. Each sampled surface is represented as the radial graph
\begin{equation}
    r(\hat{\psi}_j,\chi_k)
    =\left[(R(\hat{\psi}_j,\chi_k)-R_{\mathrm{ax}})^2
    +(Z(\hat{\psi}_j,\chi_k)-Z_{\mathrm{ax}})^2\right]^{1/2}.
\end{equation}
For a comparison geometry $c$, such as $c=\mathrm{gqdsk}$ for a G-EQDSK geometry or $c=\mathrm{ref}$ for a high-order VEQ reference, the pointwise shape difference is
\begin{equation}
    d_{r,jk}^{(c)}
    =r_{\mathrm{VEQ}}(\hat{\psi}_j,\chi_k)-r^{(c)}(\hat{\psi}_j,\chi_k).
\end{equation}
We denote retained local or profile error distributions by $\varepsilon$. The radial shape-error profile and scalar RMS shape error are
\begin{align}
    \varepsilon_{\mathrm{shape}}(\hat{\psi}_j)
     & =\left[\frac{1}{N_\chi}\sum_{k=1}^{N_\chi}\left(d_{r,jk}^{(c)}\right)^2\right]^{1/2},                \\
    E_{\mathrm{shape}}
     & =\left[\frac{1}{N_\psi}\sum_{j=1}^{N_\psi}\varepsilon_{\mathrm{shape}}(\hat{\psi}_j)^2\right]^{1/2}.
\end{align}
The degenerate $\hat{\psi}=0$ entry is evaluated as the magnetic-axis displacement. In Section~\ref{sec:representative}, $E_{\mathrm{gqdsk}}$, $E_{\mathrm{ref}}$ and $E_{\mathrm{lcfs}}$ use this same RMS construction with different comparison geometries: $E_{\mathrm{gqdsk}}$ is the direct G-EQDSK diagnostic, $E_{\mathrm{ref}}$ is the reduced-order error used for the Pareto front relative to the corresponding high-order VEQ reference, and $E_{\mathrm{lcfs}}$ is the boundary-only fit error. These $E$-metrics compare radial-graph flux-surface geometry, not a two-dimensional $\psi(R,Z)$ field error or a force-balance residual.

Unless otherwise stated, representative-case Pareto statistics use $N_\psi=11$ flux-surface entries and $N_\chi=16$ geometric-angle samples on non-degenerate surfaces. The dimensional metric has units of length; when comparing devices with different absolute sizes, the compact tables report the dimensionless ratio $E/a$, where $a$ is the minor-radius scale in the VEQ parameterization $R=R_0+a[h+\rho\cos\bar{\theta}]$. On the LCFS, the degree-one angle map $\bar{\theta}$ attains phases with $\cos\bar{\theta}=\pm1$, so $(R_{\max}^{\mathrm{lcfs}}-R_{\min}^{\mathrm{lcfs}})/2=a$.

All timings were obtained using the \texttt{VEQPy} source checkout \texttt{Zhang2026-v1} on an Intel Core i5-14600KF CPU (Ubuntu 24.04.4 LTS), with the supported Numba backend and the modified Powell hybrid method exposed through SciPy's \texttt{hybr} interface under a single-threaded CPU protocol. Each timing value is a script-recorded median over repeated solves after an untimed warm-up that triggers operator precomputation, workspace allocation and JIT compilation. The timing window excludes setup, case construction, G-EQDSK processing, plotting and exported-state generation. A minimal audit of the three Figure~\ref{fig:high-order-reconstructions} high-order cases places these one-time preprocessing, operator-setup and JIT/warm-up costs between $O(10^2)$ ms and about two seconds. These are setup costs, not hot-path costs in a post-setup repeated-query setting.

The reported values therefore describe repeated nonlinear solves after warm-up. They quantify the low-latency regime targeted here and should not be read as a matched end-to-end benchmark against complete equilibrium pipelines \cite{Lutjens1996,Lee2015}. In particular, the present implementation exports a continuous VEQ state rather than a G-EQDSK file. The timing claim is therefore a solve-only repeated-query latency claim for the fixed-boundary VEQ operator, not a replacement timing for full CHEASE-, EFIT- or ECOM-style workflows that include their own preprocessing, discretization, I/O and solver-specific convergence stages.

The Powell-hybrid solves use a projected-residual tolerance of $10^{-6}$. Pareto fronts and selected timing medians use finite-valued solver returns under the prescribed cold-start protocol, and the selected representative rows all satisfy the tolerance. The diagnostic value reported as $\varepsilon_{\mathrm{proj}}$ is the stored final Euclidean norm of the unscaled residual vector returned by the VEQ residual map,
\begin{equation}
    \varepsilon_{\mathrm{proj}}
    =\left\|\mathcal{R}(\mathbf{x}_*)\right\|_2 .
    \label{eq:projected-residual-norm}
\end{equation}
The components of $\mathcal{R}$ are the shape and profile residual equations assembled after the selected input route has been converted to the common normalized source variables. Solver wrappers may use additional block scaling for conditioning, but the tabulated $\varepsilon_{\mathrm{proj}}$ values are computed from this unscaled residual vector, not from a solver-scaled norm.

\subsection{Resolution dependence}

\begin{figure}[tb]
    \centering
    \begin{minipage}[t]{\singlecolumnwidth}
        \centering
        \includegraphics[width=\linewidth]{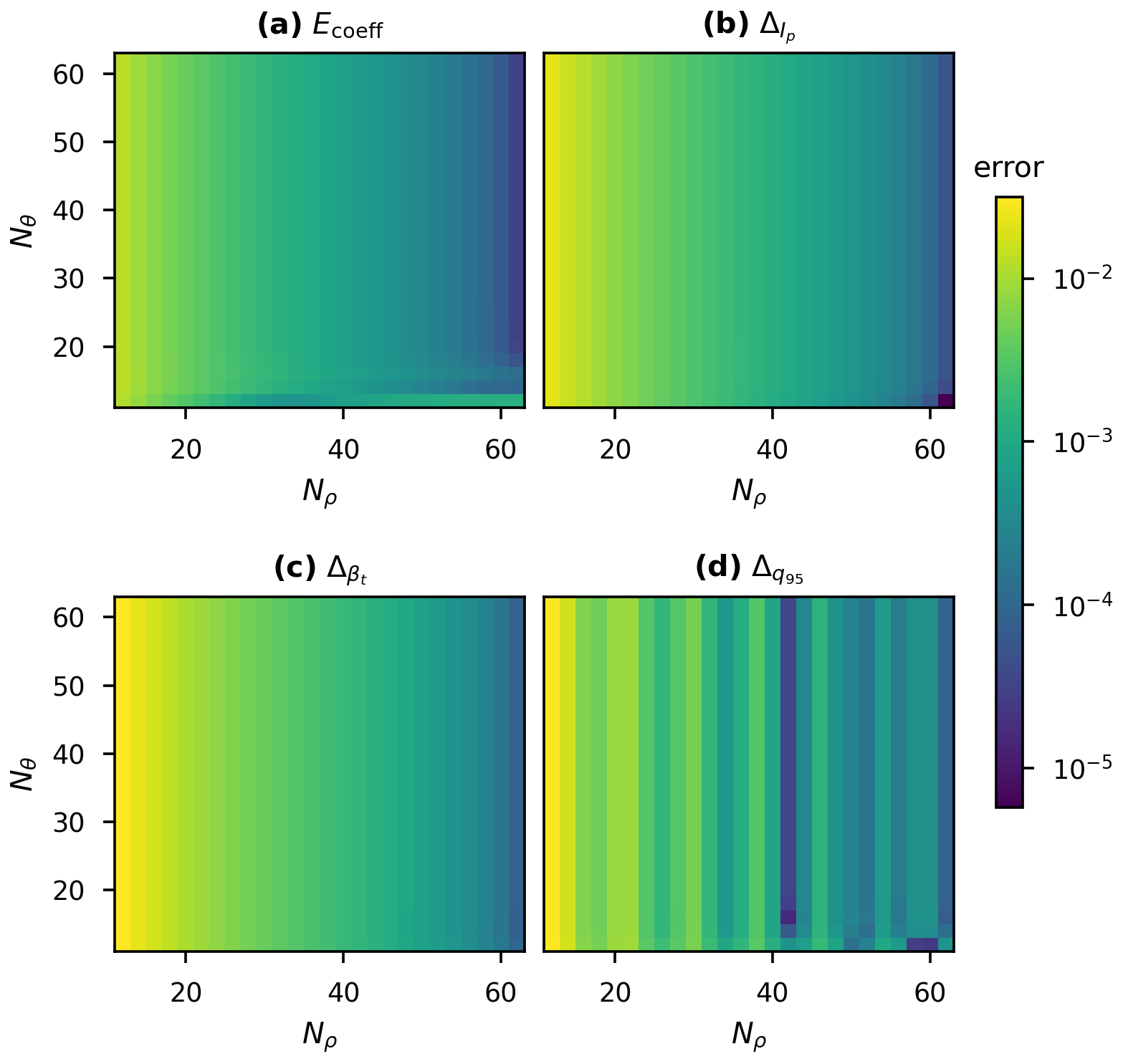}
        \caption{Grid-convergence behavior over the tested radial and poloidal resolutions. The main trend is radial-resolution controlled once the flux-surface geometry is adequately resolved, while near-edge $q_{95}$ remains more sensitive than volume-integrated quantities.}
        \label{fig:resolution-dependence}
    \end{minipage}\hfill
    \begin{minipage}[t]{\singlecolumnwidth}
        \centering
        \includegraphics[width=\linewidth]{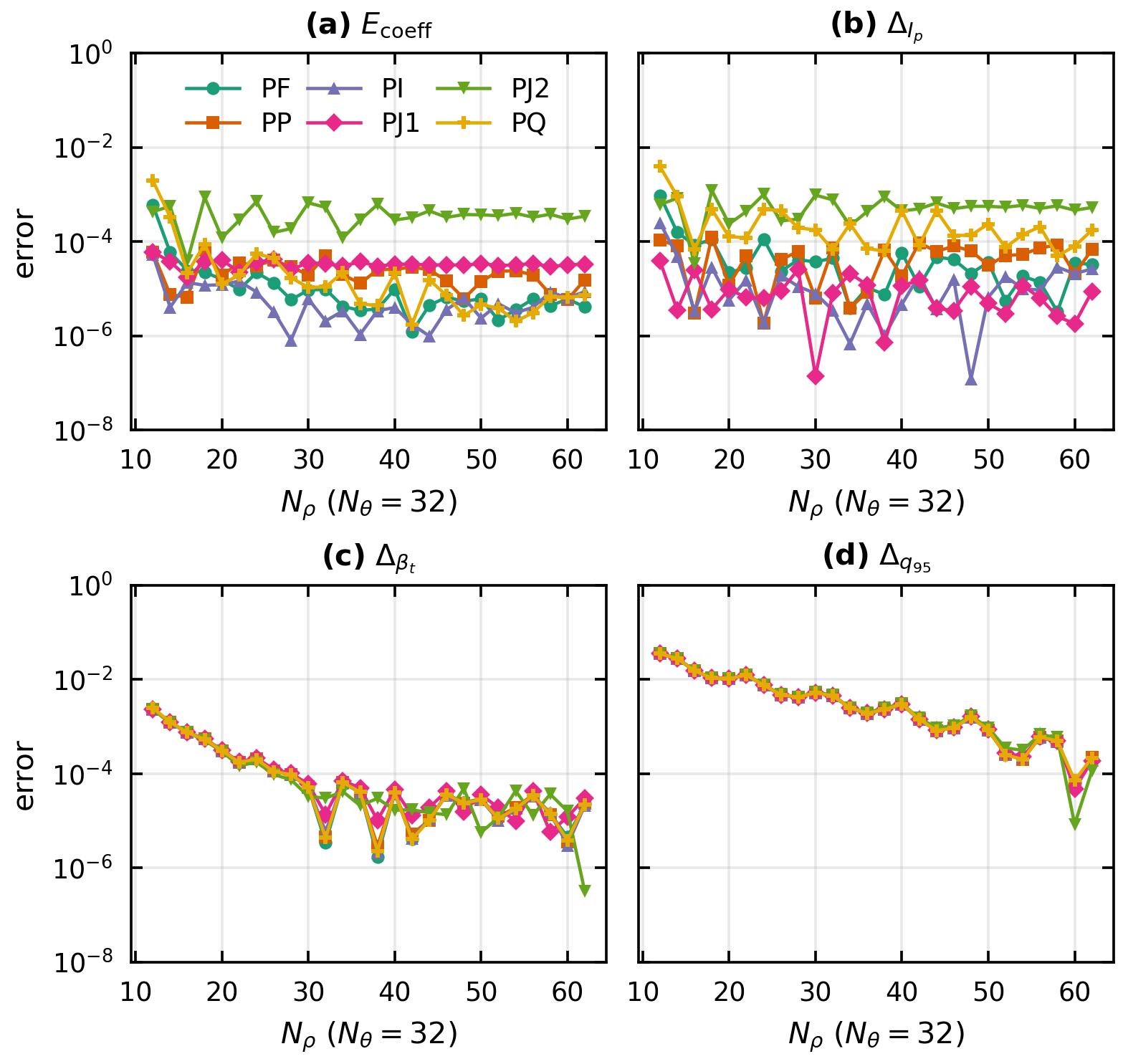}
        \caption{Consistency across input routes derived from the same controlled reference equilibrium. Different supplied quantities and closure paths converge toward the same fixed-boundary residual system for smooth mutually consistent inputs.}
        \label{fig:multi-route-consistency}
    \end{minipage}
\end{figure}

Figure~\ref{fig:resolution-dependence} reports the PF-route resolution scan. The reference solution is computed on a higher-resolution grid, while the test cases vary the radial and poloidal quadrature resolutions. The shape-coefficient, current and beta errors decrease as the quadrature is refined. Once the flux-surface geometry is adequately resolved, the dominant sensitivity is radial, consistent with the one-dimensional recovery of $I_p$, $\beta_t$ and $q$. The $q_{95}$ error is less smooth because this near-edge diagnostic depends on $\hat{\psi}_\rho$ through the safety-factor relation; small changes in radial interpolation, edge geometry or source recovery can therefore be amplified.

No degradation of projected-residual convergence was observed over the tested quadrature resolutions. Resolution statements should therefore distinguish volume-integrated quantities from near-boundary profile samples, especially for downstream tools that use edge safety factor or magnetic shear.

\subsection{Multi-route consistency}

Figure~\ref{fig:multi-route-consistency} compares the six input routes in Table~\ref{tab:input-routes} for smooth mutually consistent inputs. For each route, the supplied profiles are generated from the same controlled reference equilibrium and interpolated onto the test grid. The route-closure stage then reconstructs the canonical source and poloidal-flux-gradient quantities before the common residual assembly is evaluated. Table~\ref{tab:route-consistency} reports the corresponding numerical errors at $N_\rho=N_\theta=32$. This comparison tests route-conversion consistency under controlled input conditions; robustness to noisy, incomplete or mutually inconsistent route inputs is not assessed here.

The routes converge toward the same reference equilibrium as the radial resolution increases. The remaining differences reflect the distinct recovery paths: routes involving current density or safety factor require additional integrations, differentiations or interpolations before common residual assembly, and the parallel-current route is the most indirect because it combines magnetic geometry with the current projection. The convergence of several scalar-error curves at moderate and high radial resolution indicates that the route-closure stage recovers the same reference representation to within the tested numerical accuracy, despite using different one-dimensional inputs. Within this controlled benchmark, the route distinction changes the supplied inputs and closure relation, not the underlying fixed-boundary Grad--Shafranov residual system. The larger entries in the $\Delta_{q_{95}}$ column of Table~\ref{tab:route-consistency} are consistent with this interpretation because $q_{95}$ is a near-edge profile sample, not a global integral constraint like $I_p$ or $\beta_t$.

\begin{table}[tb]
    \caption{Route-consistency errors at $N_\rho=32$ and $N_\theta=32$ for the controlled benchmark.}
    \tableformat
    \begin{tabular}{l c c c c}
        \hline
        Route & $E_{\mathrm{coeff}}$  & $\Delta_{I_p}$        & $\Delta_{\beta_t}$    & $\Delta_{q_{95}}$     \\
        \hline
        PF    & $9.817\times 10^{-6}$ & $4.447\times 10^{-5}$ & $3.446\times 10^{-6}$ & $4.556\times 10^{-3}$ \\
        PP    & $5.049\times 10^{-5}$ & $7.254\times 10^{-5}$ & $4.631\times 10^{-6}$ & $4.561\times 10^{-3}$ \\
        PI    & $2.081\times 10^{-6}$ & $3.518\times 10^{-6}$ & $5.887\times 10^{-6}$ & $4.526\times 10^{-3}$ \\
        PJ1   & $3.396\times 10^{-5}$ & $8.127\times 10^{-6}$ & $1.377\times 10^{-5}$ & $4.559\times 10^{-3}$ \\
        PJ2   & $8.037\times 10^{-6}$ & $7.581\times 10^{-6}$ & $3.648\times 10^{-6}$ & $4.559\times 10^{-3}$ \\
        PQ    & $1.097\times 10^{-5}$ & $6.824\times 10^{-5}$ & $4.374\times 10^{-6}$ & $4.544\times 10^{-3}$ \\
        \hline
    \end{tabular}
    \label{tab:route-consistency}
\end{table}

\section{Representative fixed-boundary applications and diagnostics}
\label{sec:representative}

Using the RMS shape definition in Section~\ref{sec:error-timing}, $E_{\mathrm{gqdsk}}$ denotes the radial shape error relative to the externally supplied G-EQDSK geometry, $E_{\mathrm{ref}}$ denotes the same error relative to the corresponding high-order VEQ reconstruction, and $E_{\mathrm{lcfs}}$ denotes the RMS error of the fitted LCFS boundary. Thus $E_{\mathrm{gqdsk}}$ and $E_{\mathrm{ref}}$ measure flux-surface geometry, while $E_{\mathrm{lcfs}}$ measures only the boundary fit. Separately, $\varepsilon_{\mathrm{proj}}$ is the projected residual norm returned to the nonlinear solver, and sampled $\mathcal{G}_{\mathrm{std}}$ is the pointwise strong-form Grad--Shafranov diagnostic evaluated after the solve. The Pareto front below uses $E_{\mathrm{ref}}$ as its reduction metric; $E_{\mathrm{gqdsk}}$, $\varepsilon_{\mathrm{proj}}$ and sampled $\mathcal{G}_{\mathrm{std}}$ remain separate diagnostics of different aspects of the equilibrium representation. This section first tests high-order flux-surface reconstruction, then reduces the representation along empirical accuracy--cost fronts, evaluates the pointwise force-balance residual components left by the projected solve, and finally checks how errors in transport-geometry factors propagate through a controlled one-dimensional transport operator.

\subsection{G-EQDSK geometry reconstruction}
\label{sec:external-target-reconstruction}

We next apply the same fixed-boundary workflow to three representative equilibria, labeled D-shaped, H-mode and X-point. The D-shaped G-EQDSK file is generated from an ITER-like analytic Solov'ev solution with $R_{\mathrm{geo}}=6.20\,\mathrm{m}$ and $a=2.00\,\mathrm{m}$ \cite{Cerfon2010}. The H-mode and X-point inputs are CHEASE and EFIT G-EQDSK files, respectively \cite{Lutjens1996,Lao1985,FreeQDSKGeqdsk}. In this section, G-EQDSK is used as an equilibrium exchange format, not as a native solver state. The H-mode case tests a more radially structured reactor-relevant profile set, while the X-point case probes how far a smooth fixed-boundary flux-surface parameterization can represent a diverted-type LCFS without claiming to resolve an exact separatrix or X-point singular structure. Together, the three G-EQDSK cases form a fixed-boundary test matrix spanning smooth and strongly shaped geometries.

For reproducibility, each G-EQDSK file is converted to VEQ inputs by a deterministic preprocessing step. The H-mode and X-point file headers report rectangular $R$-$Z$ grids for poloidal flux with sizes of $128\times128$ for the CHEASE file and $257\times257$ for the EFIT file. The LCFS point sequence in each G-EQDSK file is fitted by the fixed-boundary MXH boundary coefficients, and $E_{\mathrm{lcfs}}$ reports the mismatch between the smooth fitted boundary and the original sampled LCFS. The one-dimensional G-EQDSK source profiles are used as PF$(\hat{\psi})$ inputs: the tabulated pressure-gradient and toroidal-field-function source profiles are supplied as functions of normalized poloidal flux and evaluated during the source stage using the default local-barycentric interpolation. The current, flux and sign conventions follow the COCOS-1 implementation convention stated in Section~\ref{sec:flux-coords}, and the scalar total-plasma-current constraint $I_p$ fixes the physical flux scale.

Table~\ref{tab:representative-route-inputs} records the G-EQDSK data source, selected route, physical size scales and scalar diagnostics for the representative cases. The size columns use the analytic construction for the D-shaped G-EQDSK and the G-EQDSK LCFS samples for the CHEASE and EFIT files, with $R_{\mathrm{geo}}=(R_{\max}^{\mathrm{LCFS}}+R_{\min}^{\mathrm{LCFS}})/2$ and $a=(R_{\max}^{\mathrm{LCFS}}-R_{\min}^{\mathrm{LCFS}})/2$. All three use the PF route with source profiles supplied as functions of normalized poloidal flux. The scalar solve constraint is the total plasma current $I_p$; $\beta_t$ and $q_{95}$ are reported here only as G-EQDSK diagnostics, not as additional constraints in these representative solves.

\begin{table}[tb]
    \caption{Representative-case G-EQDSK sources, route inputs, size scales and scalar diagnostic values. The signs of $I_p$ and $q_{95}$ follow the source-file current, flux and COCOS convention; their magnitudes are the relevant scalar diagnostics for the comparisons here.}
    \tableformat
    \begin{tabular}{l c l c c c c c}
        \hline
        Case     & G-EQDSK source   & Route                              & $R_{\mathrm{geo}}$ [m] & $a$ [m] & $I_p$ [MA] & $\beta_t$ & $q_{95}$ \\
        \hline
        D-shaped & Solov'ev G-EQDSK & PF($\hat{\psi}$); $I_p$ constraint & $6.20$                 & $2.00$  & $-15.00$   & $0.03$    & $-2.80$  \\
        H-mode   & CHEASE G-EQDSK   & PF($\hat{\psi}$); $I_p$ constraint & $1.00$                 & $0.65$  & $1.05$     & $0.15$    & $7.62$   \\
        X-point  & EFIT G-EQDSK     & PF($\hat{\psi}$); $I_p$ constraint & $1.66$                 & $0.62$  & $1.44$     & $0.05$    & $3.78$   \\
        \hline
    \end{tabular}
    \label{tab:representative-route-inputs}
\end{table}

\begin{figure}[tb]
    \centering
    \includegraphics[width=\doublecolumnwidth]{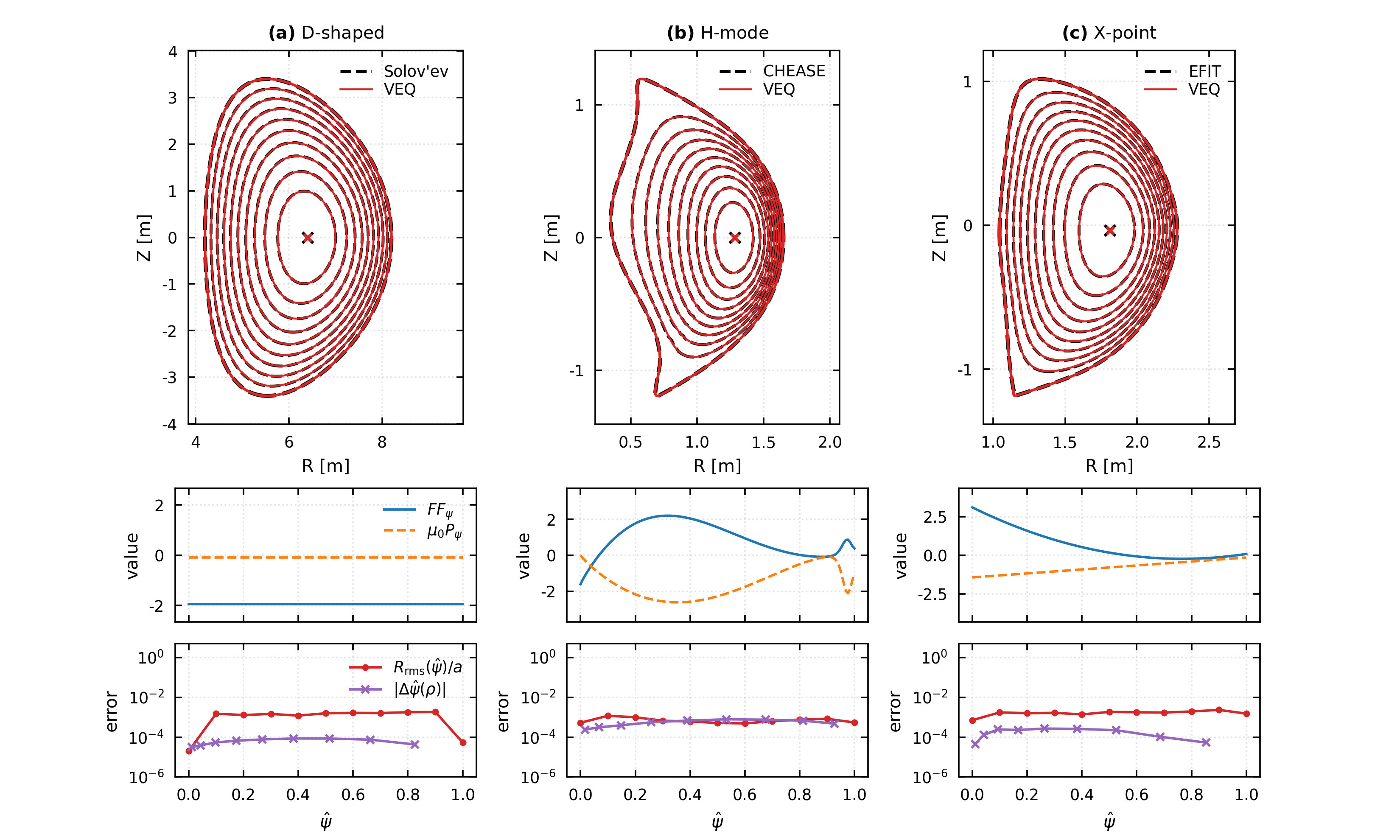}
    \caption{High-order VEQ reconstructions of the Solov'ev-, CHEASE- and EFIT-derived G-EQDSK fixed-boundary cases.}
    \label{fig:high-order-reconstructions}
\end{figure}

Figure~\ref{fig:high-order-reconstructions} compares reconstructed flux-surface geometry and radial diagnostics with the G-EQDSK geometries in Table~\ref{tab:representative-route-inputs}. This is a geometric comparison; pointwise strong-form Grad--Shafranov residuals are assessed separately in Section~\ref{sec:residual-distribution}. The lower panels show the normalized poloidal-flux mismatch and minor-radius-normalized radial-distance errors between the continuous VEQ representation and the sampled G-EQDSK data. Table~\ref{tab:high-order-validation} summarizes the high-order configurations, G-EQDSK shape errors and fixed-boundary fit errors. Here Params is the full nonlinear unknown dimension, including active $\hat{\psi}$-profile coefficients. In the count columns, Core denotes the active radial coefficient counts ordered as $(h,v,\kappa,\hat{\psi})$, while Cos lists the cosine-side shaping families $c_0,c_1,\ldots$ and Sin lists $s_1,s_2,\ldots$.

After normalization by $a$, the $E_{\mathrm{gqdsk}}$ values are $1.41\times 10^{-3}$, $7.33\times 10^{-4}$ and $1.69\times 10^{-3}$ for D-shaped, H-mode and X-point, respectively. These values include the flux-surface extraction, source-profile processing and sampling differences associated with comparing a continuous VEQ representation to a G-EQDSK exchange geometry. The LCFS-only fit errors are small but not uniformly smaller than the flux-surface reconstruction errors, because $E_{\mathrm{lcfs}}$ measures only the fitted boundary curve whereas $E_{\mathrm{gqdsk}}$ averages over sampled flux surfaces. These high-order VEQ rows become the continuous references for the reduced-representation accuracy--cost study.

The exported diagnostics are reevaluated from the supplied one-dimensional source profiles and the continuous flux-surface geometry, not fitted independently after convergence. Local Grad--Shafranov residual behavior is assessed separately in Section~\ref{sec:residual-distribution}.

\begin{table}[tb]
    \caption{High-order VEQ geometric reconstruction errors normalized by the minor-radius parameter $a$. Error symbols are defined at the start of Section~\ref{sec:representative}. Params is the full nonlinear unknown dimension, and Time is the post-setup solve-only median. Count columns list active radial coefficients; Core is ordered as $(h,v,\kappa,\hat{\psi})$, while Cos and Sin summarize the active cosine and sine shaping families.}
    \tableformat
    \begin{tabular}{l r r c c c c c}
        \hline
        Case     & Params & Time [ms] & $E_{\mathrm{gqdsk}}/a$ & $E_{\mathrm{lcfs}}/a$ & Core               & Cos                  & Sin                  \\
        \hline
        D-shaped & $75$   & $7.06$    & $1.41\times 10^{-3}$   & $1.95\times 10^{-4}$  & $(10, 0, 10, 10)$  & --                   & $(10, 5^{\times 7})$ \\
        H-mode   & $130$  & $116.13$  & $7.33\times 10^{-4}$   & $1.06\times 10^{-3}$  & $(10, 10, 10, 10)$ & $(10, 5^{\times 7})$ & $(10, 5^{\times 7})$ \\
        X-point  & $130$  & $22.90$   & $1.69\times 10^{-3}$   & $4.55\times 10^{-4}$  & $(10, 10, 10, 10)$ & $(10, 5^{\times 7})$ & $(10, 5^{\times 7})$ \\
        \hline
    \end{tabular}
    \label{tab:high-order-validation}
\end{table}

\subsection{Accuracy--cost trade-off for reduced VEQ representations}
\label{sec:pareto-reduction}

\begin{figure}[tb]
    \centering
    \includegraphics[width=\doublecolumnwidth]{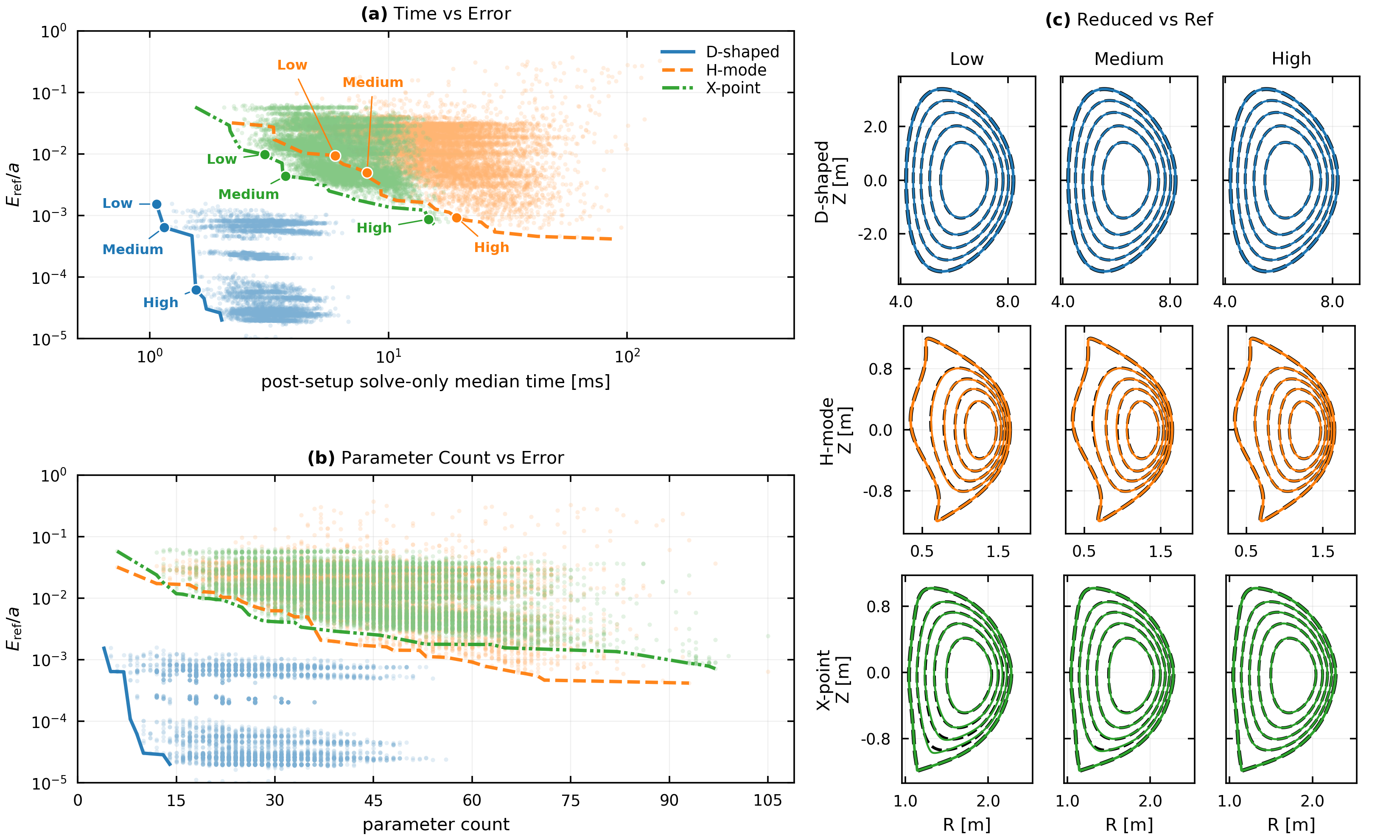}
    \caption{Accuracy--cost Pareto analysis for reduced VEQ configurations. The fronts use $E_{\mathrm{ref}}$, the reduction error relative to the high-order VEQ reconstructions; $E_{\mathrm{gqdsk}}$ is a separate direct G-EQDSK diagnostic rather than the ordering metric. The plotted fronts show empirical envelopes for the tested active-family sweep and solve protocol. Low, Medium and High denote the three representative Pareto-selected reduced configurations for each case, ordered from the lowest-latency selected row to the most accurate selected reduced row; Ref denotes the corresponding high-order VEQ reconstruction.}
    \label{fig:pareto-analysis}
\end{figure}

The preceding subsection shows that the tested high-order settings reproduce the Solov'ev-, CHEASE- and EFIT-derived G-EQDSK cases with normalized RMS radial shape errors of $1.41\times 10^{-3}$, $7.33\times 10^{-4}$ and $1.69\times 10^{-3}$. The Pareto study asks a different question: how much of that high-order VEQ representation can be retained at lower active order, and at what cost? Accordingly, the Pareto error in Figure~\ref{fig:pareto-analysis} and the $E_{\mathrm{ref}}$ column in Table~\ref{tab:pareto-configurations} are defined relative to the corresponding high-order VEQ reconstruction, not directly relative to the original G-EQDSK data. $E_{\mathrm{gqdsk}}$ remains a separate direct G-EQDSK diagnostic. The dimensionless $10^{-3}$--$10^{-2}$ range is emphasized because errors below that range may already be limited by the high-order reference itself, including contributions from G-EQDSK discretization, flux-surface sampling and boundary fitting. All timing values in this Pareto study are solve-only times and follow the protocol in Section~\ref{sec:error-timing}.

For each G-EQDSK case, three representative reduced configurations are selected along the Pareto front and are denoted Low, Medium and High in the figures below. These labels indicate increasing active representation size among the selected Pareto rows, from the lowest-latency representative point to the most accurate selected reduced point. Ref denotes the corresponding high-order VEQ reconstruction in Table~\ref{tab:high-order-validation}.

\begin{table}[tb]
    \caption{Pareto-selected reduced-order configurations. $E_{\mathrm{ref}}/a$ measures normalized reduction error relative to the high-order VEQ reference and is the Pareto-front error, while $E_{\mathrm{gqdsk}}/a$ is a separate direct G-EQDSK diagnostic. Within each case, the three rows correspond to the Low, Medium and High representative reduced configurations used in Figures~\ref{fig:pareto-analysis}, \ref{fig:residual-distribution} and \ref{fig:downstream-check}. Parentheses in the Case column give the active parameter count; count columns follow Table~\ref{tab:high-order-validation}.}
    \tableformat
    \begin{tabular}{l r c c c c c}
        \hline
        Case (Params) & Time [ms] & $E_{\mathrm{ref}}/a$ & $E_{\mathrm{gqdsk}}/a$ & Core         & Cos               & Sin               \\
        \hline
        D-shaped (4)  & $1.07$    & $1.54\times 10^{-3}$ & $2.44\times 10^{-3}$   & $(1,0,1,1)$  & --                & $(1)$             \\
        D-shaped (5)  & $1.15$    & $6.45\times 10^{-4}$ & $1.61\times 10^{-3}$   & $(1,0,2,1)$  & --                & $(1)$             \\
        D-shaped (9)  & $1.56$    & $6.24\times 10^{-5}$ & $1.39\times 10^{-3}$   & $(2,0,2,3)$  & --                & $(2)$             \\
        H-mode (27)   & $5.97$    & $9.50\times 10^{-3}$ & $9.39\times 10^{-3}$   & $(6,1,4,5)$  & $(4,1,1)$         & $(3,1,1)$         \\
        H-mode (40)   & $8.14$    & $4.98\times 10^{-3}$ & $4.82\times 10^{-3}$   & $(7,1,10,2)$ & $(5,2,2,1)$       & $(5,2,2,1)$       \\
        H-mode (65)   & $19.31$   & $9.34\times 10^{-4}$ & $1.12\times 10^{-3}$   & $(7,7,6,8)$  & $(6,4,3,2,2,1)$   & $(6,5,3,2,2,1)$   \\
        X-point (20)  & $3.03$    & $9.93\times 10^{-3}$ & $1.01\times 10^{-2}$   & $(5,2,3,4)$  & $(2,1)$           & $(2,1)$           \\
        X-point (29)  & $3.71$    & $4.41\times 10^{-3}$ & $4.48\times 10^{-3}$   & $(4,4,4,3)$  & $(2,2,2,1)$       & $(2,2,2,1)$       \\
        X-point (94)  & $14.73$   & $8.75\times 10^{-4}$ & $1.85\times 10^{-3}$   & $(8,7,9,7)$  & $(9,5,5,5,4,2,1)$ & $(9,5,5,5,5,2,1)$ \\
        \hline
    \end{tabular}
    \label{tab:pareto-configurations}
\end{table}

The dense point clouds in Figure~\ref{fig:pareto-analysis} show all tested truncation choices, while the plotted fronts retain the non-dominated configurations in $E_{\mathrm{ref}}$ and post-setup solve time. All entries in this comparison use the same $32\times32$ residual quadrature grid, so the scan isolates the effect of reducing the active representation. Grid thinning is not treated as a separate optimization axis here. The resulting Pareto front should be read as an empirical accuracy--cost envelope for the tested active families, G-EQDSK cases, initialization and solver/hardware protocol.

The sweep used the PF route with normalized poloidal-flux source-profile inputs and an $I_p$ scalar constraint. Trial signatures were generated by a pruned structured active-family sweep over core radial orders and Fourier-shaping families, capped below $10^4$ reduced candidates per case and augmented by one-step neighborhoods around the representative rows. The final tested candidate counts, excluding the high-order reference rows, are 8022 for D-shaped, 9483 for H-mode and 9478 for X-point. All three cases produced finite-valued Powell-hybrid solver returns under this protocol; if one additionally requires the final projected residual norm to be below $10^{-6}$, all D-shaped and X-point candidates pass and 9184 of 9483 H-mode candidates pass. The non-dominated fronts and all selected representative rows satisfy the projected-residual tolerance.

Within each case, Table~\ref{tab:pareto-configurations} gives representative configurations selected along the Pareto front at illustrative normalized $E_{\mathrm{ref}}/a$ levels. At the most accurate selected level, the D-shaped, H-mode and X-point configurations differ from their high-order VEQ references by $6.24\times 10^{-5}$, $9.34\times 10^{-4}$ and $8.75\times 10^{-4}$, respectively, with direct G-EQDSK shape errors of $1.39\times 10^{-3}$, $1.12\times 10^{-3}$ and $1.85\times 10^{-3}$. The corresponding active counts and post-setup solve-only medians are 9 and 1.56 ms for D-shaped, 65 and 19.31 ms for H-mode, and 94 and 14.73 ms for X-point. The fronts show that the selected D-shaped High reduced row reaches sub-$10^{-4}$ reduction error, while the selected H-mode and X-point High reduced rows reach sub-$10^{-3}$ reduction error.

Because the high-order VEQ reference is affected by boundary fitting, source-profile processing and sampling, a reduced configuration can occasionally have a smaller $E_{\mathrm{gqdsk}}$ than its $E_{\mathrm{ref}}$ ordering might suggest. This does not contradict the Pareto ordering, which is based on $E_{\mathrm{ref}}$.

\subsection{Projected residual and sampled pointwise strong-form residual diagnostics}
\label{sec:residual-distribution}

\begin{figure}[tb]
    \centering
    \includegraphics[width=\doublecolumnwidth]{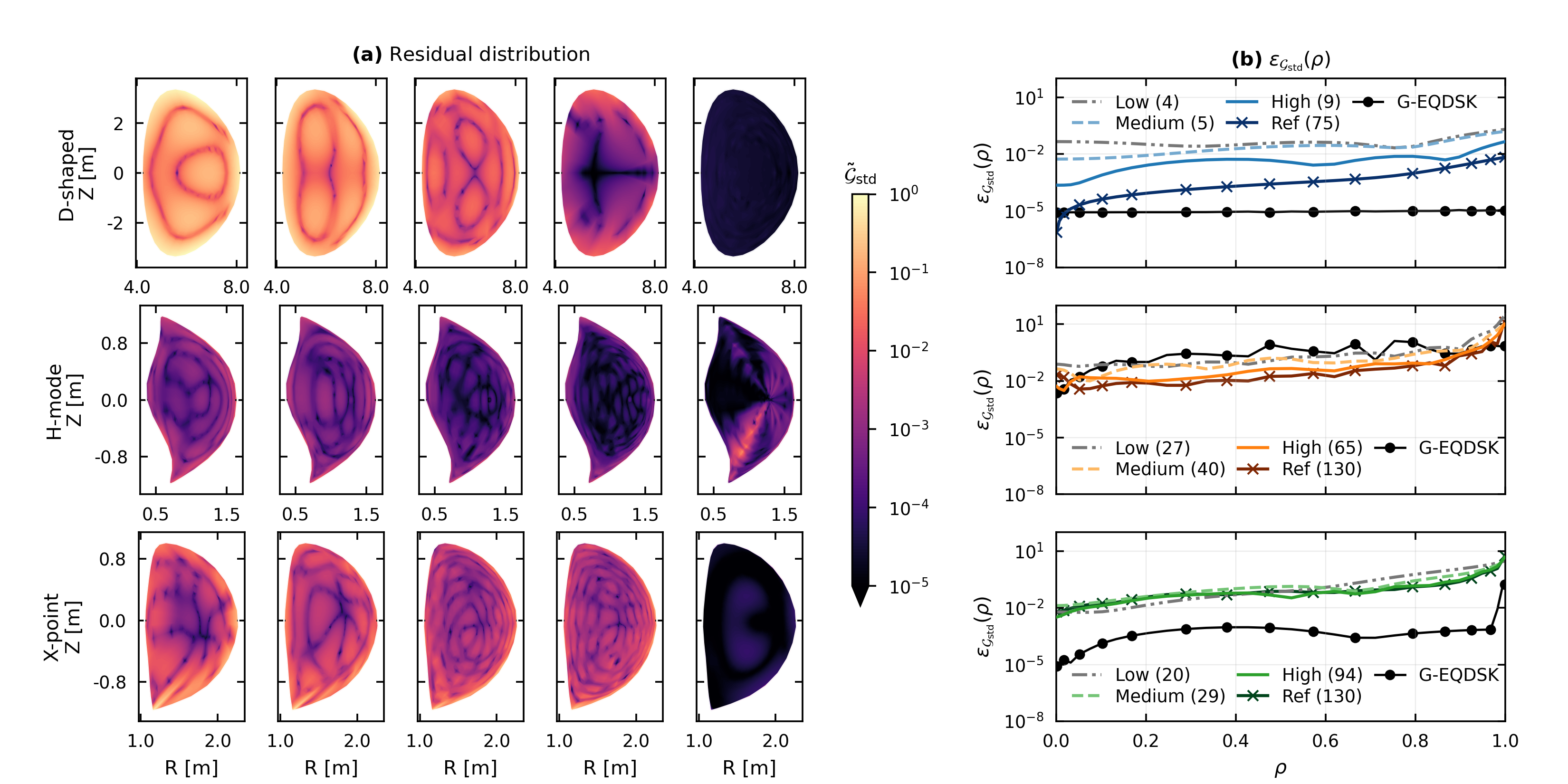}
    \caption{Sampled pointwise strong-form residual maps and radial diagnostic profiles. These maps are diagnostics of the pointwise standard-form residual component left after the projected solve, not the VEQ optimization target. Panel (a) shows case-normalized $\tilde{\mathcal{G}}_{\mathrm{std}}$ for Low, Medium, High, Ref and G-EQDSK columns, emphasizing diagnostic localization and its change with active parameter count; panel (b) shows the unnormalized poloidal RMS profile $\varepsilon_{\mathcal{G}_{\mathrm{std}}}$. Colors in panel (a) are normalized separately for each case and should not be compared across cases.}
    \label{fig:residual-distribution}
\end{figure}

We distinguish the projected residual norm $\varepsilon_{\mathrm{proj}}$, defined in Eq.~\eqref{eq:projected-residual-norm} and used as the nonlinear solve objective, from the sampled pointwise strong-form Grad--Shafranov diagnostic $\mathcal{G}_{\mathrm{std}}=(R/J)\mathcal{G}$. The former is the finite-dimensional condition enforced by the main solve; the latter is the pointwise cylindrical force-balance residual evaluated on the solve grid after convergence. It identifies spatial structure left outside the projected variational objective and should not be read as the objective minimized by VEQ or as a cross-code residual norm. Figure~\ref{fig:residual-distribution} plots $\mathcal{G}_{\mathrm{std}}$ for the Low, Medium and High reduced configurations, the high-order VEQ reference, and the G-EQDSK diagnostic. The heatmaps use the dimensionless case-local display normalization
\begin{equation}
    \tilde{\mathcal{G}}_{\mathrm{std}}(\rho,\theta)
    =
    \frac{|\mathcal{G}_{\mathrm{std}}(\rho,\theta)|}
    {\max\!\left(\{|\mathcal{G}_{\mathrm{std}}|\}_{\mathrm{case}}\right)},
\end{equation}
where the maximum is taken over all finite residual samples plotted for the same case. This normalization is purely for localization and should not be used to compare absolute residual magnitudes between cases. Panel (b) and Table~\ref{tab:residual-diagnostics} retain unnormalized standard-form residual values, including the poloidal RMS profile $\varepsilon_{\mathcal{G}_{\mathrm{std}}}(\rho)=[N_\theta^{-1}\sum_k\mathcal{G}_{\mathrm{std}}(\rho,\theta_k)^2]^{1/2}$. No universal acceptance threshold is assigned to these sampled pointwise diagnostics.

Table~\ref{tab:residual-diagnostics} reports $\mathrm{RMS}_{\mathrm{all}}(\mathcal{G}_{\mathrm{std}})$, an interior/near-edge split at $\hat{\psi}=0.8$, and $|\mathcal{G}_{\mathrm{std}}|_{\max}$. The split at $\hat{\psi}=0.8$ is a diagnostic threshold, not a pedestal, separatrix or physical-boundary definition. The G-EQDSK rows are exchange-file diagnostics under the present operator; the CHEASE and EFIT G-EQDSK rows are not native CHEASE or EFIT residuals (see Section~\ref{sec:geqdsk-interpretation}). The VEQ rows show that enrichment primarily improves the interior standard-form residual in the selected High and reference configurations, while the global RMS and maxima remain governed by near-boundary structure in the H-mode and X-point cases and do not track the Pareto shape metric monotonically in all rows.

\begin{table}[tb]
    \caption{Sampled pointwise strong-form residual diagnostics for G-EQDSK, reduced-order and high-order rows. The projected residual norm $\varepsilon_{\mathrm{proj}}$ is the nonlinear solve objective, whereas the sampled $\mathcal{G}_{\mathrm{std}}$ statistics diagnose the pointwise strong-form Grad--Shafranov residual after the solve. Values are unnormalized standard-form residual diagnostics for within-case localization and reduced-order comparison; parentheses give active parameter counts, and the RMS split at $\hat{\psi}=0.8$ is diagnostic only. G-EQDSK rows are exchange-file diagnostics; CHEASE and EFIT G-EQDSK rows are not native solver residuals.}
    \tableformat
    \begin{tabular}{l c c c c c}
        \hline
        Case             & $\varepsilon_{\mathrm{proj}}$ & $\mathrm{RMS}_{\mathrm{all}}(\mathcal{G}_{\mathrm{std}})$ & $\mathrm{RMS}_{<0.8}(\mathcal{G}_{\mathrm{std}})$ & $\mathrm{RMS}_{\geq0.8}(\mathcal{G}_{\mathrm{std}})$ & $|\mathcal{G}_{\mathrm{std}}|_{\max}$ \\
        \hline
        D-shaped (4)     & $2.10\times 10^{-8}$          & $8.18\times 10^{-2}$                                      & $3.87\times 10^{-2}$                              & $0.16$                                               & $0.34$                                \\
        D-shaped (5)     & $1.39\times 10^{-8}$          & $6.01\times 10^{-2}$                                      & $2.02\times 10^{-2}$                              & $0.12$                                               & $0.27$                                \\
        D-shaped (9)     & $6.76\times 10^{-8}$          & $1.46\times 10^{-2}$                                      & $4.07\times 10^{-3}$                              & $3.03\times 10^{-2}$                                 & $0.12$                                \\
        D-shaped (75)    & $8.02\times 10^{-9}$          & $2.39\times 10^{-3}$                                      & $5.44\times 10^{-4}$                              & $5.00\times 10^{-3}$                                 & $0.01$                                \\
        D-shaped G-EQDSK & --                            & $9.21\times 10^{-6}$                                      & $8.82\times 10^{-6}$                              & $1.05\times 10^{-5}$                                 & $1.77\times 10^{-5}$                  \\
        H-mode (27)      & $2.46\times 10^{-8}$          & $6.51$                                                    & $0.14$                                            & $11.64$                                              & $139.68$                              \\
        H-mode (40)      & $1.88\times 10^{-9}$          & $2.61$                                                    & $8.53\times 10^{-2}$                              & $4.67$                                               & $45.35$                               \\
        H-mode (65)      & $5.09\times 10^{-10}$         & $2.15$                                                    & $3.44\times 10^{-2}$                              & $3.84$                                               & $41.20$                               \\
        H-mode (130)     & $1.85\times 10^{-11}$         & $2.49$                                                    & $1.97\times 10^{-2}$                              & $4.45$                                               & $48.78$                               \\
        H-mode G-EQDSK   & --                            & $0.51$                                                    & $0.43$                                            & $0.65$                                               & $7.28$                                \\
        X-point (20)     & $2.97\times 10^{-8}$          & $1.03$                                                    & $0.29$                                            & $2.13$                                               & $14.22$                               \\
        X-point (29)     & $1.97\times 10^{-8}$          & $0.81$                                                    & $0.15$                                            & $1.71$                                               & $13.52$                               \\
        X-point (94)     & $3.91\times 10^{-9}$          & $1.06$                                                    & $7.74\times 10^{-2}$                              & $2.25$                                               & $20.04$                               \\
        X-point (130)    & $3.32\times 10^{-10}$         & $1.14$                                                    & $7.14\times 10^{-2}$                              & $2.44$                                               & $19.78$                               \\
        X-point G-EQDSK  & --                            & $3.52\times 10^{-2}$                                      & $5.17\times 10^{-4}$                              & $7.52\times 10^{-2}$                                 & $0.54$                                \\
        \hline
    \end{tabular}
    \label{tab:residual-diagnostics}
\end{table}

The numerical split makes the localization pattern explicit. For H-mode, the interior statistic falls from 0.14 at 27 parameters to $3.44\times10^{-2}$ at 65 parameters, and the 130-parameter high-order reference reaches $1.97\times10^{-2}$. The corresponding near-edge statistic decreases from 11.64 to 3.84 for the selected rows but remains large at high order. For X-point, the interior RMS drops from 0.29 at 20 parameters to $7.74\times10^{-2}$ at 94 parameters and $7.14\times10^{-2}$ at high order, but the near-edge statistic remains around 1.7--2.4 and the global RMS is not reduced monotonically. Thus the main improvement in the H-mode and X-point cases is interior force-balance reduction rather than near-edge cancellation; in the D-shaped data, relative to the lowest-order selected row, the high-order reference row reduces both interior and near-edge statistics by more than an order of magnitude. This behavior is consistent with the finite-dimensional projection in Eq.~\eqref{eq:finite-projected-residual}: once the projected residual is small, the remaining sampled standard-form component is the part not controlled by the selected ansatz-induced test functions and route-closure moments.

Likely contributors to the observed near-boundary localization include metric differentiation in $\Delta^*\psi$, source-profile closure and the fixed-boundary ansatz: interior corrections vanish at $\rho=1$, so boundary-fit mismatch cannot in general be fully compensated by interior shape variations alone. The source routes reduce inputs to one-dimensional flux-function closures such as $P_\psi(\rho)$, $FF_\psi(\rho)$ and $\psi_\rho(\rho)$; for a fitted two-dimensional geometry, these radial closures cannot absorb arbitrary near-boundary poloidal force-balance components introduced by boundary fitting, file-grid interpolation or the finite active representation. For solver-exported G-EQDSK files such as CHEASE and EFIT, the exchange file and native solver state are distinct objects, and the exported rectangular grid must be differentiated and resampled before evaluating $\mathcal{G}_{\mathrm{std}}$. The maps screen localization and downstream applicability; they do not require pointwise vanishing of the standard-form residual.

Increasing the active representation improves the interior diagnostic most clearly in the selected High and reference rows and reduces the D-shaped standard-form residual in absolute magnitude. In the H-mode and X-point cases, however, the remaining signal is largely concentrated toward the boundary. For boundary-sensitive downstream quantities, the geometry metrics, projected residual norm and pointwise maps should therefore be interpreted together. They identify radial regions where the reduced representation is adequate and cases that may require boundary treatment, local correction or higher-fidelity follow-up.

Appendix~\ref{app:collocation-comparison} compares these projected variational states with an unanchored point-collocation least-squares polish on the same finite VEQ representation. The polish can reduce some sampled pointwise strong-form residual components, but it solves a different finite-dimensional objective, redistributes residuals in radius and adds cost. It is therefore retained as a diagnostic comparison rather than as the main solve.

\subsection{Scope-controlled 1-D transport-geometry coupling test}
\label{sec:downstream-coupling}

\begin{figure}[tb]
    \centering
    \includegraphics[width=\doublecolumnwidth]{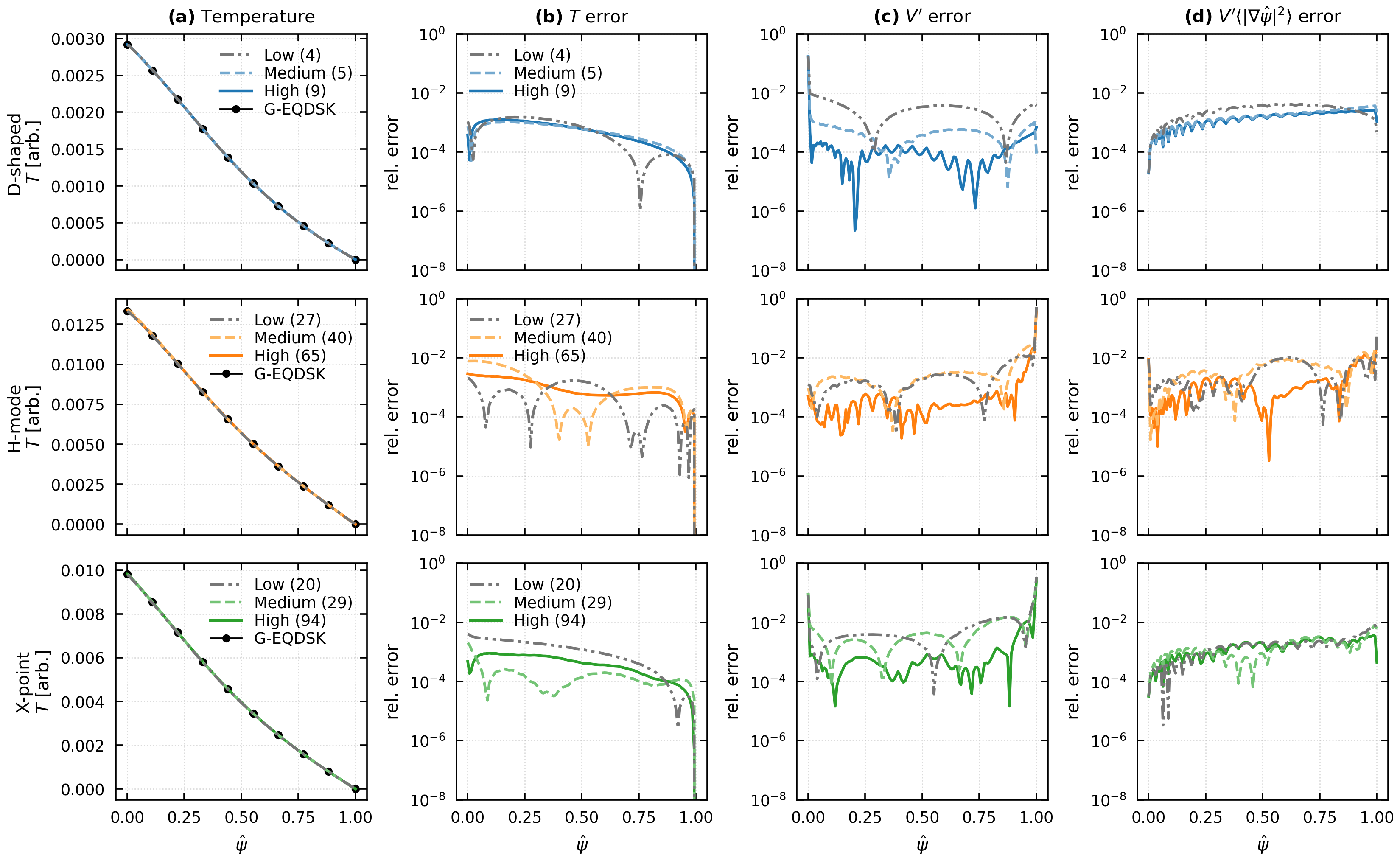}
    \caption{Scope-controlled one-dimensional transport-geometry coupling test for the D-shaped, H-mode and X-point reduced configurations against the target geometry read from G-EQDSK. Parentheses in the Low, Medium and High legend labels give the active parameter counts.}
    \label{fig:downstream-check}
\end{figure}

The residual maps identify boundary-sensitive regions, while many integrated-modeling components use only flux-surface geometric factors over a prescribed radial domain. We therefore use an isolated downstream coupling test rather than a full integrated-workflow simulation: only the VEQ-dependent geometry channel is varied. Specifically, errors in $V'$ and $V'\langle|\nabla\hat{\psi}|^2\rangle$ are propagated through a prescribed one-dimensional flux-coordinate heat-diffusion operator, where $V'=\mathrm{d}V/\mathrm{d}\hat{\psi}$ is evaluated on the normalized poloidal-flux coordinate. The Low, Medium and High reduced configurations defined in Section~\ref{sec:pareto-reduction} are passed to this operator on the same normalized poloidal-flux coordinate. For each case, $T(\hat{\psi})$ solves
\begin{equation}
    -\frac{\mathrm{d}}{\mathrm{d}\hat{\psi}}
    \left[
        V'\langle|\nabla\hat{\psi}|^2\rangle
        \frac{\mathrm{d}T}{\mathrm{d}\hat{\psi}}
        \right]
    = V'S(\hat{\psi}),\quad T(1)=0,\quad
    \left.\frac{\mathrm{d}T}{\mathrm{d}\hat{\psi}}\right|_{\hat{\psi}=0}
    =0,
\end{equation}
corresponding to unit diffusivity in this normalized test. The source shape is a representative smooth test profile chosen to exercise the geometry factors in the one-dimensional operator, not an experimental heating fit or an additional validation target:
\begin{equation*}
    S(\hat{\psi})=C
    \exp\left[-\left(\frac{\hat{\psi}-0.18}{0.33}\right)^2\right]
    \left(1-0.15\hat{\psi}\right),\quad \int_0^1 V'_{\mathrm{tgt}}S\,\mathrm{d}\hat{\psi}=1.
\end{equation*}
Here $V'_{\mathrm{tgt}}$ denotes the target geometry factor read from G-EQDSK. After this normalization, the same $S(\hat{\psi})$ is held fixed for all reduced configurations; only the geometry factors in the operator are changed. The total stored-energy diagnostic in Table~\ref{tab:downstream-check} is
\begin{equation*}
    W=\frac{3}{2}\int_0^1 V'(\hat{\psi})T(\hat{\psi})\,\mathrm{d}\hat{\psi}.
\end{equation*}
The toroidal-beta proxy treats the solved temperature profile as a pressure-like surrogate with the volume-average convention in Section~\ref{sec:input-routes}; its relative error is reported as $\Delta_{\beta_t}$. All errors below are measured against the corresponding target read from G-EQDSK curve shown in Figure~\ref{fig:downstream-check}. In the relative-error panels,
$\delta_f(\hat{\psi})=|f(\hat{\psi})-f_{\mathrm{tgt}}(\hat{\psi})|/\max_{\hat{\psi}}|f_{\mathrm{tgt}}(\hat{\psi})|$;
Table~\ref{tab:downstream-check} reports the RMS of these relative-error profiles.

\begin{table}[tb]
    \caption{Downstream relative profile RMS errors and scalar relative errors against the target read from G-EQDSK for the scope-controlled transport-geometry coupling test in Figure~\ref{fig:downstream-check}.}
    \tableformat
    \begin{tabular}{l c c c c c}
        \hline
        Case         & \(\mathrm{RMS}(\delta_T)\) & \(\mathrm{RMS}(\delta_{V'})\) & \(\mathrm{RMS}(\delta_{V'\langle|\nabla\hat{\psi}|^2\rangle})\) & \(\Delta_W\)         & \(\Delta_{\beta_t}\) \\
        \hline
        D-shaped (4) & $8.69\times 10^{-4}$       & $1.63\times 10^{-2}$          & $2.95\times 10^{-3}$                                            & $2.03\times 10^{-3}$ & $1.20\times 10^{-3}$ \\
        D-shaped (5) & $6.54\times 10^{-4}$       & $1.51\times 10^{-2}$          & $1.95\times 10^{-3}$                                            & $8.27\times 10^{-4}$ & $1.71\times 10^{-5}$ \\
        D-shaped (9) & $7.36\times 10^{-4}$       & $1.50\times 10^{-2}$          & $1.66\times 10^{-3}$                                            & $4.08\times 10^{-4}$ & $4.37\times 10^{-4}$ \\
        H-mode (27)  & $8.57\times 10^{-4}$       & $4.64\times 10^{-2}$          & $7.77\times 10^{-3}$                                            & $5.19\times 10^{-4}$ & $6.72\times 10^{-3}$ \\
        H-mode (40)  & $3.14\times 10^{-3}$       & $4.68\times 10^{-2}$          & $5.90\times 10^{-3}$                                            & $6.14\times 10^{-3}$ & $1.26\times 10^{-2}$ \\
        H-mode (65)  & $1.34\times 10^{-3}$       & $3.84\times 10^{-2}$          & $4.26\times 10^{-3}$                                            & $2.89\times 10^{-3}$ & $8.72\times 10^{-3}$ \\
        X-point (20) & $1.68\times 10^{-3}$       & $3.30\times 10^{-2}$          & $2.31\times 10^{-3}$                                            & $6.61\times 10^{-3}$ & $4.22\times 10^{-3}$ \\
        X-point (29) & $3.18\times 10^{-4}$       & $3.25\times 10^{-2}$          & $2.04\times 10^{-3}$                                            & $2.26\times 10^{-3}$ & $1.33\times 10^{-4}$ \\
        X-point (94) & $5.36\times 10^{-4}$       & $2.61\times 10^{-2}$          & $1.85\times 10^{-3}$                                            & $8.36\times 10^{-4}$ & $1.84\times 10^{-3}$ \\
        \hline
    \end{tabular}
    \label{tab:downstream-check}
\end{table}

Across the selected reduced configurations, the temperature-profile diagnostics remain below about one percent and the stored-energy and beta-proxy diagnostics have sub-percent to percent-level errors for this fixed-source problem, even when the profiles of transport-geometry factors show larger RMS differences. Because this test isolates only the geometry channel, boundary-sensitive applications should still retain the pointwise residual and geometry screening discussed above.

\section{Discussion}
\label{sec:discussion}

\subsection{Scope of the fixed-boundary validation}

The benchmarks support the present implementation as a fast fixed-boundary parametric Grad--Shafranov solver and as a repeated-query provider of equilibrium geometry for modeling workflows. Because the nonlinear solve, source-profile normalization, diagnostics and resampling are all evaluated from the same continuous nested-surface representation, the converged VEQ state can be reused directly by downstream models rather than treated as a solver-specific rectangular-grid output or as a purely analytic shape closure. In this sense, the present implementation shares the repeated-query objective of integrated-modeling tools such as METIS \cite{Artaud2018}, while solving the narrower fixed-boundary task of supplying flux-surface geometry, profile normalization, current and safety-factor diagnostics, and transport-geometry factors after the LCFS and source-profile inputs have been specified.

Section~\ref{sec:downstream-coupling} should be read in this restricted geometry-channel sense. The one-dimensional test propagates errors in $V'$ and $V'\langle|\nabla\hat{\psi}|^2\rangle$ through a prescribed flux-coordinate heat-diffusion operator while holding the source shape and transport model fixed. It therefore probes the sensitivity of a representative downstream calculation to VEQ-supplied transport-geometry factors; it is not a validation of a complete integrated-modeling workflow with transport closures, source modeling, boundary evolution or platform-specific coupling choices.

The residual diagnostics define the main applicability boundary of the compact finite-dimensional representation. A small projected value means that the Grad--Shafranov residual has been reduced in the test directions induced by the active VEQ manifold and the route-dependent profile closures; it does not imply pointwise cancellation of the standard-form force-balance residual. The near-boundary localization observed in the H-mode and X-point cases is therefore consistent with the selected trial and test spaces, fixed-boundary treatment, source closure, differentiation and quadrature. When local force-balance accuracy is the primary requirement, especially for boundary-sensitive, separatrix-sensitive or stability-sensitive studies, solvers or correction stages with direct pointwise or collocation residual control are the more appropriate reference class. VEQ makes this distinction explicit by reporting both projected and sampled pointwise diagnostics.

Accordingly, the present validation supports geometry-dominated repeated fixed-boundary calls, not all equilibrium use cases. Coil currents, passive structures, separatrix determination and plasma--vacuum coupling are outside the benchmark. The X-point case is treated as a smoothed fixed-boundary G-EQDSK geometry extracted from exchange data, rather than as an exact separatrix or free-boundary equilibrium.

\subsection{Relation to point-collocation solvers and adjacent equilibrium workflows}
\label{sec:desc-relation}

A direct comparison with DESC or other point-collocation solvers would require a matched target definition, boundary treatment, source convention, residual operator and timing protocol. DESC is a pseudo-spectral equilibrium solver that evaluates nonlinear force-balance equations on collocation grids for variables represented in a solver-native spectral basis \cite{Dudt2020}. VEQ instead solves a square finite-dimensional system of variational and moment residuals generated by the MXH--Chebyshev flux-surface ansatz and by normalized one-dimensional closures associated with the selected input route. The two formulations therefore differ in unknowns, residual definitions, weighting, boundary treatment, dimensionality and convergence certificates. An unmatched head-to-head timing or residual comparison would conflate representation choice, route construction, diagnostic processing and implementation details, rather than isolate the parametric representation and projected residual solve proposed here.

Within that scope, the validation is organized around the intended fixed-boundary use case: route consistency for mutually compatible inputs, minor-radius-normalized flux-surface reconstruction against Solov'ev-, CHEASE- and EFIT-derived G-EQDSK cases, post-setup repeated-solve cost and sampled pointwise strong-form residual diagnostics. The sampled $\mathcal{G}_{\mathrm{std}}$ maps provide the relevant pointwise force-balance screen for the VEQ outputs, while the projected residual reports convergence of the finite-dimensional equation actually solved. Appendix~\ref{app:collocation-comparison} is therefore a within-representation comparison on the same VEQ finite manifold, not a cross-code benchmark.

The resulting comparisons are staged rather than head-to-head. The route tests check closure consistency under mutually compatible inputs; the G-EQDSK comparisons test flux-surface geometry recovered from exchange data; the reduced-versus-high-order VEQ comparisons test representation compression; Appendix~\ref{app:collocation-comparison} compares finite-dimensional residual objectives on the same manifold; and the transport calculation tests propagation through a prescribed geometry channel. None of these comparisons is presented as a native-residual or end-to-end runtime ranking of CHEASE, EFIT, ECOM, DESC, METIS or Miller-type tools.

The same caution applies to adjacent equilibrium workflows, although the relevant distinction changes by tool class. CHEASE, EFIT, ECOM and DESC occupy the high-fidelity or solver-native equilibrium and reconstruction side of the spectrum, with their own variables, boundary treatments and convergence certificates. METIS and Miller-type closures are rapid reduced-modeling tools with intentionally compressed geometry. The present realization is intended for an intermediate fixed-boundary role: it retains more flux-surface geometry and residual diagnostics than analytic engineering closures, while remaining lighter and more reusable in repeated fixed-boundary calls than a full equilibrium or reconstruction pipeline. This positioning is why the paper reports source normalization conventions, post-setup solve latency, pointwise residual screening and downstream propagation of transport-geometry errors, rather than a single cross-code speed or residual ranking.

\subsection{Interpretation of G-EQDSK data}
\label{sec:geqdsk-interpretation}

The representative external geometries are supplied through G-EQDSK files, as described in Section~\ref{sec:external-target-reconstruction}: the D-shaped case is generated from the analytic Solov'ev solution, while the H-mode and X-point cases are CHEASE and EFIT exports \cite{Lutjens1996,Lao1985,FreeQDSKGeqdsk}. These files are exchange representations, not native solver states. Differences are therefore expected because rectangular G-EQDSK poloidal-flux grids, fitted LCFS boundaries and MXH--Chebyshev flux-surface representations are distinct objects. Interpolation, flux-surface extraction, metric reconstruction and COCOS conventions become part of the effective comparison workflow, especially near X-points or weak-poloidal-field regions \cite{Sauter2013,Kripner2019PLEQUE,Garcia2022INGRID}. Related constrained fitting work makes the same distinction between flux-surface parameterization and reference-grid data \cite{Snoep2023}.

The residual rows computed from G-EQDSK data have the same exchange-data interpretation. CHEASE and EFIT satisfy the corresponding Grad--Shafranov balance in their own discrete formats, boundary treatments, source definitions and solution variables; their native residuals are not contained in the exported G-EQDSK files. A residual reconstructed from G-EQDSK---for example by finite differences on the exported $(R,Z)$ grid followed by interpolation to the VEQ flux-coordinate nodes as in Table~\ref{tab:residual-diagnostics}---therefore tests consistency of the exchange data under the chosen diagnostic operator, not the native solver residual of the original CHEASE or EFIT calculation.

\subsection{Limitations and future extensions}

The reported errors and solve-only timings depend on the chosen active families, backend, solver configuration and hardware. The diagnostics separate projected variational control from sampled pointwise standard-form residual control, and the present implementation is axisymmetric and fixed-boundary. Stability-sensitive, separatrix-sensitive or edge force-balance-sensitive applications should therefore retain pointwise residual screening and, when needed, use local correction stages or higher-fidelity solvers with direct pointwise residual control. Extending VEQ implementations to free-boundary problems will require coupling to vacuum fields, external currents and boundary variations \cite{Amorisco2024}. Extending them to three-dimensional equilibria would require a corresponding magnetic-surface representation and three-dimensional metric and residual formulation, and should be assessed against three-dimensional equilibrium solvers only under matched problem definitions and diagnostics \cite{Hirshman1983,Dudt2020}.

A natural next application is coupling VEQ to 0-D/1-D systems, transport or source models as a geometry-rich equilibrium closure. In that role, the solver would provide low-order modeling components with continuous flux-surface geometry, normalized source-profile closure and low-cost repeated solves, while selected boundary-sensitive cases could still be passed to reconstruction, free-boundary or pointwise-residual-controlled tools.

The corresponding technical extensions are boundary-residual control for stability-sensitive downstream applications, coupling to free-boundary or three-dimensional constraints, and assessment of VEQ as a warm start for higher-fidelity equilibrium solvers.

\section{Conclusion}
\label{sec:conclusion}

This work presents VEQ as a parametric equilibrium framework and evaluates its present fixed-boundary realization as a fast Grad--Shafranov solver for continuous fixed-boundary tokamak geometry. The formulation uses an MXH--Chebyshev flux-surface representation and maps the PF, PP, PI, PJ1, PJ2 and PQ input routes into a common finite-dimensional residual assembly. The main solve reported here enforces a variationally induced projected residual, while sampled pointwise strong-form diagnostics and the optional point-collocation least-squares polish examine residual components outside the finite VEQ test space. Controlled benchmarks show route consistency for smooth mutually compatible profiles generated from the same reference equilibrium, with the largest scalar differences occurring in the near-edge $q_{95}$ diagnostic rather than in global integral quantities. The representative G-EQDSK applications use the PF route; the remaining routes are tested here only in the controlled mutually compatible setting, leaving noisy or inconsistent non-PF route stress tests to future work.

For the D-shaped, H-mode and X-point tests, the high-order VEQ reconstructions reach normalized direct RMS radial G-EQDSK shape errors $E_{\mathrm{gqdsk}}/a$ of $1.41\times 10^{-3}$, $7.33\times 10^{-4}$ and $1.69\times 10^{-3}$ against the Solov'ev-, CHEASE- and EFIT-derived G-EQDSK cases. The most accurate Pareto-selected reduced configurations then retain normalized reduction errors $E_{\mathrm{ref}}/a$ of $6.24\times 10^{-5}$, $9.34\times 10^{-4}$ and $8.75\times 10^{-4}$ relative to those high-order references, with direct G-EQDSK shape errors of $1.39\times 10^{-3}$, $1.12\times 10^{-3}$ and $1.85\times 10^{-3}$ and post-setup solve-only median times of 1.56, 19.31 and 14.73 ms. These timings correspond to the repeated-solve protocol defined above, excluding G-EQDSK processing, setup, JIT compilation and export-state construction. They demonstrate millisecond-scale fixed-boundary VEQ solve latency, not matched end-to-end replacement timings for CHEASE-, EFIT- or ECOM-style workflows.

Sampled pointwise strong-form diagnostics identify an important applicability boundary of the present formulation. A converged VEQ solution satisfies a finite projected system of the form $\mathcal{A}^{T}\mathcal{W}\mathbf{g}=0$, not the pointwise condition $\mathbf{g}=0$. Nonzero sampled $\mathcal{G}_{\mathrm{std}}$ maps are therefore expected for a finite representation unless the exact equilibrium and all numerical representations are mutually compatible with the chosen ansatz. Enriching the active representation reduces the interior standard-form Grad--Shafranov force-balance residual most clearly. In the H-mode and X-point cases, the remaining signal is concentrated near the boundary, consistent with finite active representation, boundary fitting, source closure and exchange-file processing. These mechanisms are not separated into unique contributions here.

Boundary-sensitive studies should therefore use the reported pointwise residual diagnostics to screen cases, or couple the VEQ baseline representation with higher-fidelity local correction. The downstream transport test should be read as a geometry-channel sensitivity check rather than as a validation of a complete transport scenario. Within this deliberately isolated setting, the Pareto-selected reduced geometries keep the temperature-profile response below about one percent and produce sub-percent to percent-level scalar-output changes for the tested cases, while RMS errors in selected transport-geometry factors remain at several percent. For geometry-dominated repeated calls, the present implementation provides a compact solved representation, provided that cases requiring stricter local force balance are identified and handled separately. Within the present fixed-boundary scope, VEQ retains two-dimensional flux-surface geometry in a form that remains inexpensive to solve and reuse.

\section*{CRediT authorship contribution statement}
\textbf{Ruohan Zhang}: Methodology, Software, Validation, Formal analysis, Investigation, Data curation, Visualization, Writing -- original draft, Writing -- review \& editing;
\textbf{Huasheng Xie}: Conceptualization, Methodology, Supervision, Writing -- review \& editing;
\textbf{Yueyan Li}: Conceptualization, Methodology;
\textbf{Weiqi Meng}: Resources (provision of G-EQDSK data used for benchmarking);
\textbf{Feng Wang}: Supervision, Writing -- review \& editing;
\textbf{Zhengxiong Wang}: Supervision, Funding acquisition.

\section*{Declaration of competing interest}
The \texttt{VEQPy} implementation used in this study is released as open-source software under the BSD-3-Clause license. The authors declare that they have no other known competing financial interests or personal relationships that could have appeared to influence the work reported in this paper.

\section*{Acknowledgements}
The authors acknowledge the open-source scientific Python community, whose software ecosystem supported the numerical analysis and visualization reported in this work. This work was supported by the National Magnetic Confinement Fusion Energy R\&D Program of China (Grant No. 2022YFE03090000) and the National Natural Science Foundation of China (Grant No. 12405265).

\section*{Data availability}
The \texttt{VEQPy} source code (\url{https://github.com/zhangtakeda/veqpy}) used for this study is available under the BSD-3-Clause license. The results reported here were generated with article-specific source checkout \texttt{Zhang2026-v1} and dependency versions recorded in the project metadata. During peer review, these article-specific artifacts are provided to editors and reviewers through an anonymized review package associated with this submission. The package includes the source checkout used for the reported runs, figure and benchmark scripts, shared configuration files, processed G-EQDSK inputs or their generation scripts, and dependency metadata. The article-specific artifacts will be released publicly after acceptance.

\appendix
\section{Route-closure formulas for canonical residual variables}
\appendixlabel{app:route-closures}

With the normalized source, profile and scale conventions introduced in Section~\ref{sec:input-routes}, every input route is converted to the canonical residual variables $(\hat{P}_\psi,\hat{FF}_\psi,\hat{\psi}_\rho)$. The formulas below specify only the route-dependent one-dimensional closures; they do not introduce different equilibrium equations. After the flux-surface-averaged balance is divided by the source scale $\alpha_1$, the common scale-explicit relation used by the closures is
\begin{equation}
    \frac{\alpha_2}{\alpha_1}
    \frac{\mathrm{d}}{\mathrm{d}\rho}
    \left(\hat{K}\hat{\psi}_\rho\right)
    +\hat{L}_\rho\hat{FF}_\psi
    +\frac{V_\rho}{4\pi^2}\hat{P}_\psi=0,
    \label{eq:pf-average-closure}
\end{equation}
with $\hat{K}$, $\hat{L}_\rho$ and $V_\rho$ evaluated from the current geometry during residual evaluation. In this appendix, ``absorbing the scale factor'' denotes the following source-closure normalization. If a route first reconstructs only an unnormalized poloidal-flux-gradient shape $u(\rho)$, the implementation sets
\begin{equation}
    \hat{\psi}_\rho(\rho)
    =\frac{u(\rho)}{\displaystyle\int_0^1u(\rho')\,\mathrm{d}\rho'}
    \quad\text{so that}\quad
    \int_0^1\hat{\psi}_\rho\,\mathrm{d}\rho=1 ,
\end{equation}
and stores the removed amplitude as the constant relation between the physical source and flux scales. The residual evaluation therefore uses the normalized $\hat{\psi}_\rho$, while the final $\alpha_1,\alpha_2$ are fixed by the allowed scalar constraint or by the direct physical-input normalization for unconstrained physical-profile inputs. Routes with an independently supplied $\hat{\psi}_\rho$ retain the factor $\alpha_2/\alpha_1$ explicitly in the missing-source closure.

\routeclosure{PF route}
This route supplies pressure-gradient and toroidal-field-function source profiles and recovers the normalized poloidal-flux-gradient profile. The $\rho$-source expression below is an intermediate closure form; after the source-coordinate conversion, the common residual receives the canonical flux-derivative profiles. For sources tabulated against the radial label $\rho$, the radial- and flux-derivative forms are related by the source-coordinate convention in Section~\ref{sec:input-routes}. Writing $Q=\hat{K}\hat{\psi}_\rho$ and imposing $Q(0)=0$, the scale-explicit equation gives $\hat{\psi}_\rho=\sqrt{\alpha_1/\alpha_2}\,X(\rho)$ with
\begin{equation*}
    X(\rho)=
    \frac{1}{\hat{K}(\rho)}
    \left[-2\int_0^\rho\hat{K}(\rho')
        \left(\hat{L}_\rho(\rho')\hat{FF}_\rho(\rho')
        +\frac{V_\rho(\rho')}{4\pi^2}\hat{P}_\rho(\rho')
        \right)\,\mathrm{d}\rho'\right]^{1/2},
\end{equation*}
followed by normalization $\hat{\psi}_\rho=X/\int_0^1X\,\mathrm{d}\rho$ and integration to recover $\hat{\psi}$. Equivalently, the removed amplitude is stored as $\alpha_2/\alpha_1=(\int_0^1X\,\mathrm{d}\rho)^2$ before the current or beta constraint fixes the remaining physical scale. For PF sources supplied as functions of normalized poloidal flux, the corresponding unnormalized profile is
\begin{equation*}
    Y(\rho)
    =-\frac{1}{\hat{K}(\rho)}\int_0^\rho\left[
        \hat{L}_\rho(\rho')\hat{FF}_\psi\bigl(\hat{\psi}(\rho')\bigr)
        +\frac{V_\rho(\rho')}{4\pi^2}
        \hat{P}_\psi\bigl(\hat{\psi}(\rho')\bigr)
        \right] \,\mathrm{d}\rho',
\end{equation*}
with $\hat{\psi}_\rho=Y/\int_0^1Y\,\mathrm{d}\rho$ and $\alpha_2/\alpha_1=\int_0^1Y\,\mathrm{d}\rho$. The PF($\hat{\psi}$) dependence remains part of the nonlinear residual because the tabulated source values are sampled on the iterate-dependent map $\hat{\psi}(\rho)$. When $\hat{\psi}$ is part of the unknown profile set, this is not a separate source-local fixed-point solve: the residual evaluation samples $P_\psi(\hat{\psi}(\rho))$ and $FF_\psi(\hat{\psi}(\rho))$ on the iterate-dependent profile, updates the scale relation above, and the moment equations associated with $\hat{\psi}$ close the flux map together with the shape equations.

\routeclosure{PQ route}
This route supplies a pressure-gradient source and the safety-factor profile. The safety-factor relation
\begin{equation}
    \hat{\psi}_\rho
    =\frac{F\hat{L}_\rho}{\alpha_2 q}
    \label{eq:pq-route-closure}
\end{equation}
expresses the normalized poloidal-flux-gradient profile through the toroidal-field function. Substituting Eq.~\eqref{eq:pq-route-closure} and
$\hat{FF}_\psi=F F_\rho/(\alpha_1\alpha_2\hat{\psi}_\rho)$ into Eq.~\eqref{eq:pf-average-closure}, and absorbing the constant source scale as in PF, gives the PQ balance
\begin{equation}
    \frac{\mathrm{d}}{\mathrm{d}\rho}
    \left(\frac{\hat{K}\hat{L}_\rho}{q}F\right)
    +q\frac{\mathrm{d}F}{\mathrm{d}\rho}
    +\frac{V_\rho}{4\pi^2}\hat{P}_\psi=0 .
    \label{eq:pq-expanded-balance}
\end{equation}
For pressure input tabulated in normalized poloidal flux, the source samples are evaluated on $\hat{\psi}(\rho)$ at each nonlinear iterate, and Eq.~\eqref{eq:pq-expanded-balance} is equivalently
\begin{equation*}
    \left(\frac{\hat{K}\hat{L}_\rho}{q}+q\right)
    \frac{\mathrm{d}F}{\mathrm{d}\rho}
    +\frac{\mathrm{d}}{\mathrm{d}\rho}
    \left(\frac{\hat{K}\hat{L}_\rho}{q}\right)F
    =-\frac{V_\rho}{4\pi^2}\hat{P}_\psi,
    \quad F(1)=R_0B_0 .
\end{equation*}
For radial-derivative pressure input, writing $Y=F^2$ gives
\begin{equation*}
    \left(\frac{\hat{K}\hat{L}_\rho}{q}+q\right)
    \frac{\mathrm{d}Y}{\mathrm{d}\rho}
    +2\frac{\mathrm{d}}{\mathrm{d}\rho}
    \left(\frac{\hat{K}\hat{L}_\rho}{q}\right)Y
    =-\frac{V_\rho \hat{P}_\rho q}{2\pi^2\hat{L}_\rho},
    \quad Y(1)=(R_0B_0)^2 .
\end{equation*}
Then $F=\operatorname{sgn}(R_0B_0)\sqrt{Y}$, Eq.~\eqref{eq:pq-route-closure} gives $\hat{\psi}_\rho$, and the exported toroidal-field source is
\begin{equation}
    \hat{FF}_\psi
    =\frac{F}{\alpha_1\alpha_2\hat{\psi}_\rho}
    \frac{\mathrm{d}F}{\mathrm{d}\rho}
    =\frac{1}{2\alpha_1\alpha_2\hat{\psi}_\rho}
    \frac{\mathrm{d}Y}{\mathrm{d}\rho} .
\end{equation}
The scale factors are then fixed by the same integral constraints used by the other routes. Radial-derivative source-profile inputs use the source-coordinate conversion in Section~\ref{sec:input-routes}; after this normalization all routes call the same transformed residual operator.

\routeclosure{PP route}
This route supplies the pressure-gradient source and normalized poloidal-flux-gradient profile. The missing toroidal-field-function source is obtained from the rearranged averaged balance,
\begin{equation*}
    \hat{FF}_\psi
    =-\frac{1}{\hat{L}_\rho}
    \left[
        \frac{\alpha_2}{\alpha_1}
        \frac{\mathrm{d}}{\mathrm{d}\rho}
        \left(\hat{K}\hat{\psi}_\rho\right)
        +\frac{V_\rho}{4\pi^2}\hat{P}_\psi
        \right].
\end{equation*}

\routeclosure{PI route}
This route supplies the enclosed toroidal current profile $I_{\mathrm{tor}}(\rho)$ directly. Using the current relation in Section~\ref{sec:input-routes}, the canonical residual variables are
\begin{align}
    \hat{\psi}_\rho
     & =\frac{\mu_0 I_{\mathrm{tor}}}{2\pi\alpha_2\hat{K}}, \\
    \hat{FF}_\psi
     & =-\frac{1}{\hat{L}_\rho}
    \left[
    \frac{\mu_0}{2\pi\alpha_1}
    \frac{\mathrm{d}I_{\mathrm{tor}}}{\mathrm{d}\rho}
    +\frac{V_\rho}{4\pi^2}\hat{P}_\psi
    \right].
    \label{eq:current-route-closure}
\end{align}
Dimensionless current inputs are scaled to the physical current convention before applying these relations.

\routeclosure{PJ1 route}
This route supplies a flux-surface-averaged toroidal-current-density shape. It is first converted to the enclosed toroidal current
\begin{equation*}
    I_{\mathrm{tor}}(\rho)
    =\int_0^\rho j_{\mathrm{tor}}(\rho')S_{\rho'}\,\mathrm{d}\rho',
\end{equation*}
and then uses the PI closure in Eq.~\eqref{eq:current-route-closure} with this reconstructed $I_{\mathrm{tor}}$.

\routeclosure{PJ2 route}
This parallel-current-density route uses active $F$-profile coefficients and the source scaling in Section~\ref{sec:input-routes} to recover
\begin{equation*}
    \hat{FF}_\rho=\frac{FF_\rho}{\alpha_1\alpha_2},
    \quad
    \hat{FF}_\psi=\frac{\hat{FF}_\rho}{\hat{\psi}_\rho} .
\end{equation*}
It obtains $\hat{\psi}_\rho$ from the supplied parallel-current-density shape,
\begin{equation*}
    I_{\mathrm{tor}}(\rho)
    =2\pi F(\rho)\int_0^\rho
    \frac{\hat{L}_\rho(\rho')j_\parallel(\rho')}{F(\rho')}\,\mathrm{d}\rho', \quad \hat{\psi}_\rho
    =\frac{\mu_0 I_{\mathrm{tor}}}{2\pi\alpha_2\hat{K}} .
\end{equation*}

\routeclosure{Scale constraints}
The normalized pressure profile is reconstructed from the selected normalized pressure-gradient source by
\begin{equation}
    \hat{P}(\rho)=-\int_\rho^1 \hat{P}_\rho(\rho')\,\mathrm{d}\rho'
    =-\int_\rho^1 \hat{P}_\psi(\rho')\hat{\psi}_\rho(\rho')\,\mathrm{d}\rho',\quad \hat{P}(1)=0.
\end{equation}
When $\hat{\psi}_\rho$ is independently reconstructed, the plasma-current and toroidal-beta constraints give
\begin{equation*}
    \alpha_2
    =\frac{\mu_0 I_p}{2\pi\hat{K}(1)\hat{\psi}_\rho(1)},
    \quad
    \alpha_1\alpha_2
    =\frac{\beta_tB_0^2}{2\langle\hat{P}\rangle_V},
\end{equation*}
where $\langle\hat{P}\rangle_V$ uses the volume-average convention defined in Section~\ref{sec:input-routes}. Routes without an independent $\hat{\psi}_\rho$ reconstruction use the corresponding one-scale constraint with $\alpha_2=\mathcal{C}_\psi\alpha_1$; unsupported constraint combinations are rejected during route setup.

\section{Point-collocation polish on the finite VEQ manifold}
\appendixlabel{app:collocation-comparison}

\begin{figure}[tb]
    \centering
    \includegraphics[width=\singlecolumnwidth]{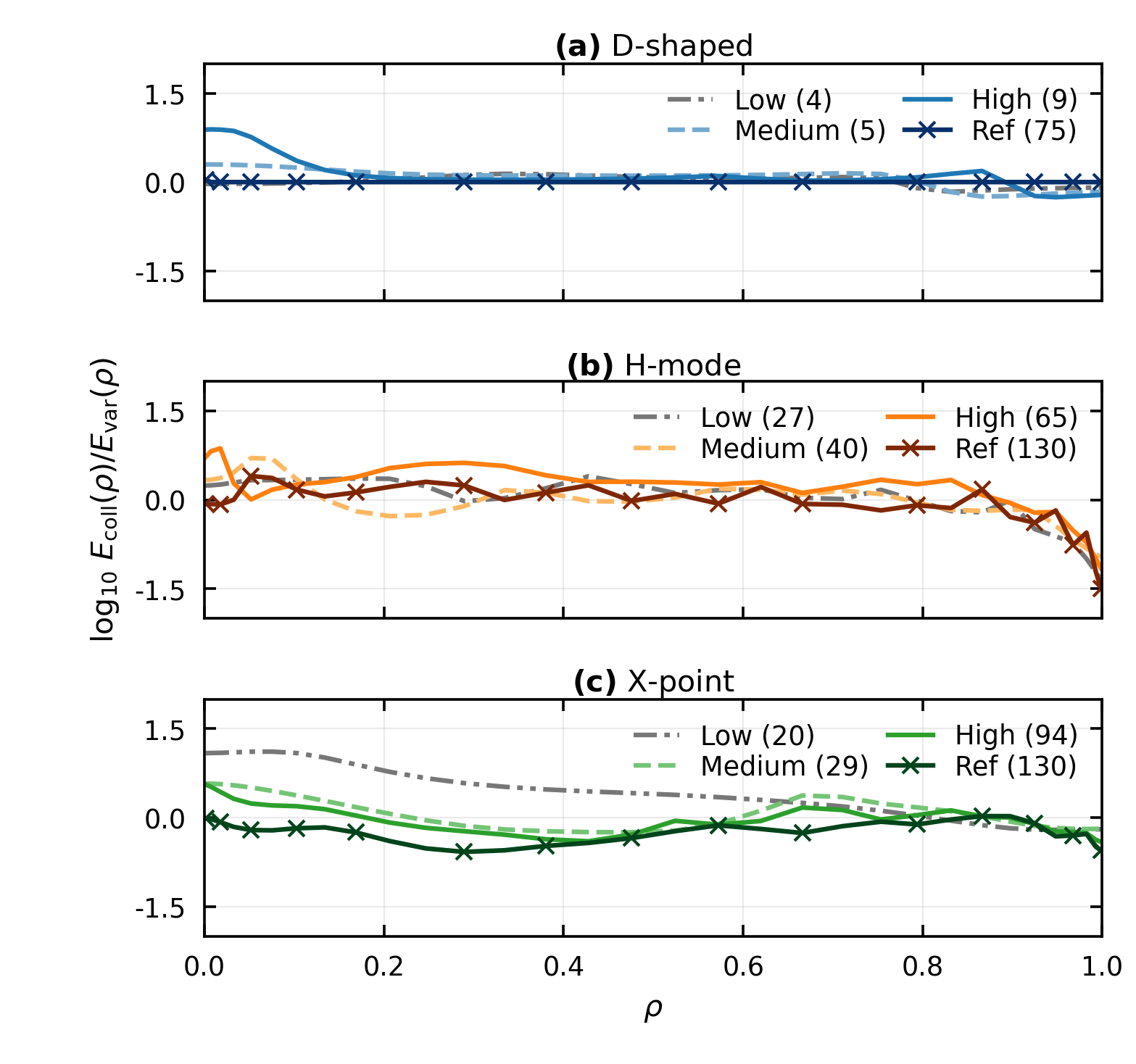}
    \caption{Radial redistribution of the sampled standard-form Grad--Shafranov residual after an unanchored point-collocation least-squares polish. Each curve shows $\log_{10}[E_{\mathrm{coll}}(\rho)/E_{\mathrm{var}}(\rho)]$, where $E(\rho)$ is the poloidal RMS of the quadrature-scaled residual used by the collocation objective. Values below zero favor the polish at that radius; values above zero favor the projected variational state. The curves are not identically zero because the two solves impose different finite-dimensional residual conditions on the same VEQ manifold.}
    \label{fig:appendix-collocation-comparison}
\end{figure}

Figure~\ref{fig:appendix-collocation-comparison} and Table~\ref{tab:appendix-collocation-ratios} compare the projected variational states used in the main text with an unanchored point-collocation least-squares polish initialized from each converged state; in the implementation this corresponds to setting the collocation weight to $\lambda=1$. This is a within-representation diagnostic, not a criterion for accepting or rejecting the projected solve: the fixed boundary, active MXH--Chebyshev space, normalized source-profile inputs and initial state are unchanged. Only the finite-dimensional residual objective is changed. In the present implementation, the collocation vector consists of quadrature-scaled samples proportional to $(R/J)\mathcal{G}$, i.e. to the strong-form residual in standard cylindrical form $\mathcal{G}_{\mathrm{std}}$ in Eq.~\eqref{eq:standard-gs-residual}, rather than to the transformed density $\mathcal{G}$ itself. The comparison identifies where the projected and collocation residual conditions disagree and how that disagreement changes with active order.

Table~\ref{tab:appendix-collocation-ratios} gives complementary integrated diagnostics for the same within-representation comparison, using a pure Levenberg--Marquardt polish on the same $32\times32$ residual grid. The geometry column reports
\[
    r_{E_{\mathrm{gqdsk}}}
    =
    E_{\mathrm{gqdsk}}(\mathbf{x}_c)/
    E_{\mathrm{gqdsk}}(\mathbf{x}_v),
\]
where $\mathbf{x}_v$ and $\mathbf{x}_c$ are the projected variational and collocation-polished coefficient vectors. This uses the same direct G-EQDSK radial-graph shape metric defined in Section~\ref{sec:error-timing}: the magnetic-axis displacement plus non-degenerate surfaces at the fixed normalized poloidal-flux entries $\hat{\psi}=0.1,\ldots,1.0$, sampled at 16 geometric angles. The VEQ surfaces are reconstructed from the continuous representation on a uniform diagnostic $(\rho,\theta)$ grid of at least $128\times256$ points before interpolation to these $\hat{\psi}$ levels. Thus the value is a fixed shape diagnostic, separate from both the $32\times32$ residual grid and the finite unknown dimension; changing the chosen diagnostic flux-surface levels or angular sampling would change the reported absolute metric, but the ratios below use the same metric for both states. The timing ratio adds the warm-start collocation polish to the already converged projected solve, so it measures diagnostic overhead rather than replacement solve-only timing.

\begin{table}[tb]
    \caption{Point-collocation polish relative to the projected variational solution on the same finite VEQ manifold. The polish minimizes quadrature-scaled samples of $\mathcal{G}_{\mathrm{std}}=(R/J)\mathcal{G}$, and the table reports collocation/variational ratios for sampled standard-form Grad--Shafranov residual diagnostics together with the direct G-EQDSK shape-error ratio $r_{E_{\mathrm{gqdsk}}}$. Values below one indicate reduction of the corresponding residual or shape diagnostic. The RMS split is at $\hat{\psi}=0.8$, and $t_{\mathrm{rel}}=(t_{\mathrm{var}}+t_{\mathrm{coll}})/t_{\mathrm{var}}$.}
    \tableformat
    \begin{tabular}{l c c c c c c}
        \hline
        Case (Params) & $r_{\mathrm{RMS,all}}$ & $r_{\mathrm{RMS},<0.8}$ & $r_{\mathrm{RMS},\geq0.8}$ & $r_{|\mathcal{G}_{\mathrm{std}}|_{\max}}$ & $r_{E_{\mathrm{gqdsk}}}$ & $t_{\mathrm{rel}}$ \\
        \hline
        D-shaped (4)  & $0.843$                & $1.026$                 & $0.799$                    & $0.744$                                   & $1.265$                  & $2.674$            \\
        D-shaped (5)  & $0.723$                & $1.146$                 & $0.668$                    & $0.541$                                   & $1.227$                  & $2.831$            \\
        D-shaped (9)  & $0.650$                & $1.207$                 & $0.597$                    & $0.389$                                   & $1.043$                  & $3.140$            \\
        D-shaped (75) & $0.999$                & $1.001$                 & $0.999$                    & $1.007$                                   & $1.000$                  & $4.055$            \\
        H-mode (27)   & $0.071$                & $1.415$                 & $0.067$                    & $0.026$                                   & $3.532$                  & $9.870$            \\
        H-mode (40)   & $0.139$                & $1.227$                 & $0.135$                    & $0.074$                                   & $3.710$                  & $10.778$           \\
        H-mode (65)   & $0.109$                & $1.978$                 & $0.105$                    & $0.037$                                   & $8.549$                  & $8.352$            \\
        H-mode (130)  & $0.050$                & $0.932$                 & $0.050$                    & $0.025$                                   & $3.341$                  & $2.796$            \\
        X-point (20)  & $0.668$                & $1.045$                 & $0.635$                    & $0.647$                                   & $4.022$                  & $7.725$            \\
        X-point (29)  & $0.677$                & $1.247$                 & $0.655$                    & $0.634$                                   & $5.461$                  & $9.750$            \\
        X-point (94)  & $0.433$                & $1.062$                 & $0.428$                    & $0.444$                                   & $6.438$                  & $16.677$           \\
        X-point (130) & $0.319$                & $0.812$                 & $0.316$                    & $0.374$                                   & $4.425$                  & $15.886$           \\
        \hline
    \end{tabular}
    \label{tab:appendix-collocation-ratios}
\end{table}

Figure~\ref{fig:appendix-collocation-comparison} and Table~\ref{tab:appendix-collocation-ratios} show that the projected variational and collocation-polished states are generally not identical. The polish redistributes the sampled standard-form Grad--Shafranov residual in radius: negative log-ratio regions in Fig.~\ref{fig:appendix-collocation-comparison} improve, regions near zero are comparable, and positive regions favor the projected state. The table gives the corresponding integrated standard-form residual diagnostics and direct G-EQDSK shape diagnostic. The polish reduces the global sampled RMS and maximum residual in every row, especially for the H-mode and X-point cases. Interior ratios above one in several reduced rows are therefore not a contradiction; they show that the global quadrature-weighted $\mathcal{G}_{\mathrm{std}}$ objective can trade a modest interior increase for a larger edge reduction, while other rows improve in both regions.

The added $r_{E_{\mathrm{gqdsk}}}$ column shows the corresponding geometric tradeoff: the D-shaped high-order row remains nearly unchanged, but most reduced rows, and especially the H-mode and X-point cases, move farther from the direct G-EQDSK flux-surface geometry after the unanchored polish. These trends are within-representation diagnostics, not a monotonic convergence guarantee for either finite-dimensional residual condition. The larger timing ratios reflect the added overdetermined least-squares polish after the projected solve has already converged.

At the continuous level there is no contradiction between weak and pointwise formulations: a state satisfying the strong-form equation in standard cylindrical form,
\begin{equation*}
    \mathcal{G}_{\mathrm{std}}(\rho,\theta;\mathbf{x}^{\dagger})=0,
    \quad \text{or equivalently}\quad
    \mathcal{G}(\rho,\theta;\mathbf{x}^{\dagger})=0
    \quad \text{for all }(\rho,\theta),
\end{equation*}
makes every weak test of $\mathcal{G}$ vanish; conversely, a sufficiently rich weak test space recovers the strong equation in the usual weak sense. The distinction appears only after restricting the equilibrium to a finite nonlinear manifold parameterized by $\mathbf{x}$, which collects the active flux-surface coefficients and route-specific profile coefficients. On a quadrature or diagnostic grid with $N_\rho N_\theta$ samples, write
\begin{equation*}
    \mathbf{g}(\mathbf{x})
    =
    \bigl[\mathcal{G}(\rho_i,\theta_j;\mathbf{x})\bigr]_{q=1}^{N_\rho N_\theta},
    \quad
    \mathbf{g}_{\mathrm{std}}(\mathbf{x})
    =
    \bigl[\mathcal{G}_{\mathrm{std}}(\rho_i,\theta_j;\mathbf{x})\bigr]_{q=1}^{N_\rho N_\theta},
    \qquad q=(i,j).
\end{equation*}
Here $\mathcal{G}_{\mathrm{std}}=(R/J)\mathcal{G}$ pointwise, but the collocation residual is best viewed as the sampled standard-form residual itself, not as a separately introduced diagonal conversion of $\mathbf{g}$.

The projected VEQ solve and the unanchored polish impose different finite-dimensional optimality conditions. The variational solve derived in Section~\ref{sec:formulation} has the algebraic form
\begin{equation}
    \mathcal{A}(\mathbf{x})^{T}\mathcal{W}\mathbf{g}(\mathbf{x})=0,
    \label{eq:app-projected-system}
\end{equation}
which is a finite Galerkin orthogonality condition, not a sampled pointwise equation. The columns of $\mathcal{A}$ are the admissible VEQ test directions: sampled convective flux variations $\chi_k=\nabla\psi\cdot\partial\mathbf{r}/\partial p_k$ for shape coefficients and finite route-closure moment tests for profile coefficients. Thus Eq.~\eqref{eq:app-projected-system} enforces orthogonality to the reduced test directions; it does not require all entries of $\mathbf{g}$ to vanish. Since typically $N_\rho N_\theta\gg n$ and $\mathrm{rank}(\mathcal{A})\le n$, nonzero sampled components can remain in the null space of $\mathcal{A}^{T}\mathcal{W}$. The pointwise $\mathcal{G}_{\mathrm{std}}$ maps in the main text visualize the corresponding standard-form force-balance diagnostic for this finite-dimensional remainder left by the trial space, test space, source closure, quadrature and boundary representation.

The collocation polish instead selects the coefficient vector by reducing the weighted sampled standard-form residual norm,
\begin{equation*}
    \mathbf{x}_{c}\in\arg\min_{\mathbf{x}}
    \frac{1}{2}\|\mathcal{W}^{1/2}\mathbf{g}_{\mathrm{std}}(\mathbf{x})\|_2^2,
\end{equation*}
using the same square-root quadrature weights in $\mathcal{W}^{1/2}$. It therefore satisfies the least-squares normal equation
\begin{equation}
    J_c^{T}\mathcal{W}^{1/2}\mathbf{g}_{\mathrm{std}}(\mathbf{x}_{c})=0, \quad
    J_c=
    \left.        \frac{\partial\!\left(\mathcal{W}^{1/2}\mathbf{g}_{\mathrm{std}}(\mathbf{x})\right)}
    {\partial\mathbf{x}}
    \right|_{\mathbf{x}=\mathbf{x}_c}.
    \label{eq:app-collocation-normal}
\end{equation}
Equation~\eqref{eq:app-collocation-normal} is therefore a sensitivity-weighted normal equation for the sampled standard-form residual, rather than Eq.~\eqref{eq:app-projected-system}. Since $\mathcal{G}_{\mathrm{std}}=(R/J)\mathcal{G}$ pointwise, its Jacobian also carries the metric-density sensitivities of this standard-form quantity. The two states coincide only when the finite representation can already make these samples essentially zero, or when the collocation sensitivity space reproduces the variational test space with the same weighting and residual density.

Equivalently, linearizing the collocation objective about the variational state shows that the unanchored polish can move whenever the remaining sampled standard-form residual is visible to the collocation sensitivity space; the variational condition $\mathcal{A}(\mathbf{x}_{v})^{T}\mathcal{W}\mathbf{g}(\mathbf{x}_{v})=0$ alone does not make this gradient vanish.

The fixed-boundary constraint makes this distinction visible. Interior shape corrections vanish at the prescribed LCFS; schematically,
\begin{equation*}
    \frac{\partial\mathbf{r}}{\partial p_k}=O(1-\rho^2),
    \qquad
    \chi_k=\nabla\psi\cdot\frac{\partial\mathbf{r}}{\partial p_k}=O(1-\rho^2)
    \quad (\rho\to1) .
\end{equation*}
Near-edge force-balance components caused by boundary fitting, metric differentiation, source closure or limited active order are therefore weakly visible to the variational shape equations, because the admissible boundary motion is constrained. A point-collocation norm still counts those edge samples directly. The unanchored polish can consequently move interior coefficients to reduce the sampled norm even when that motion is not the flux-surface-geometry correction selected by the projected variational form. Because the coefficient-to-residual and metric maps are nonlinear, a smaller sampled residual norm is not by itself a smaller geometry error:
\begin{equation*}
    \|\mathcal{W}^{1/2}\mathbf{g}_{\mathrm{std}}(\mathbf{x}_1)\|_2
    <
    \|\mathcal{W}^{1/2}\mathbf{g}_{\mathrm{std}}(\mathbf{x}_2)\|_2
    \quad\not\Rightarrow\quad
    E_{\mathrm{shape}}(\mathbf{x}_1)<E_{\mathrm{shape}}(\mathbf{x}_2).
\end{equation*}

This comparison does not rank collocation methods. It shows only that, for the present fixed-boundary VEQ manifold, the projected variational solve and an unanchored sampled least-squares polish are different finite-dimensional problems. The main solve is kept variational because its test directions are generated by the same constrained finite manifold that represents the equilibrium; sampled pointwise strong-form residuals are retained as diagnostics rather than as the defining objective for moving the geometry.

\bibliographystyle{elsarticle-num-names}
\bibliography{references}

@article{Shafranov1966,
  author  = {Shafranov, V. D.},
  title   = {Plasma Equilibrium in a Magnetic Field},
  journal = {Reviews of Plasma Physics},
  volume  = {2},
  pages   = {103},
  year    = {1966},
  note    = {Edited by M. A. Leontovich; Consultants Bureau, New York}
}

@inproceedings{GradRubin1958,
  author    = {Grad, H. and Rubin, H.},
  title     = {Hydromagnetic Equilibria and Force-Free Fields},
  booktitle = {Proceedings of the Second United Nations International Conference on the Peaceful Uses of Atomic Energy},
  volume    = {31},
  pages     = {190},
  address   = {Geneva},
  year      = {1958}
}

@article{Kruskal1958,
  title     = {Equilibrium of a Magnetically Confined Plasma in a Toroid},
  volume    = {1},
  issn      = {0031-9171},
  doi       = {10.1063/1.1705884},
  number    = {4},
  journal   = {The Physics of Fluids},
  publisher = {AIP Publishing},
  author    = {Kruskal, M. D. and Kulsrud, R. M.},
  year      = {1958},
  pages     = {265--274}
}

@article{Greene1971,
  title     = {Tokamak Equilibrium},
  volume    = {14},
  issn      = {0031-9171},
  doi       = {10.1063/1.1693488},
  number    = {3},
  journal   = {The Physics of Fluids},
  publisher = {AIP Publishing},
  author    = {Greene, John M. and Johnson, John L. and Weimer, Katherine E.},
  year      = {1971},
  pages     = {671--683}
}

@article{Palha2016,
  author  = {Palha, Artur and Koren, Barry and Felici, Federico},
  title   = {A mimetic spectral element solver for the {Grad--Shafranov} equation},
  journal = {Journal of Computational Physics},
  volume  = {316},
  pages   = {63--93},
  year    = {2016},
  doi     = {10.1016/j.jcp.2016.04.002}
}

@article{SanchezVizuet2019,
  author  = {S{\'a}nchez-Vizuet, Tonatiuh and Solano, Manuel E.},
  title   = {A Hybridizable Discontinuous Galerkin solver for the {Grad--Shafranov} equation},
  journal = {Computer Physics Communications},
  volume  = {235},
  pages   = {120--132},
  year    = {2019},
  doi     = {10.1016/j.cpc.2018.09.013}
}

@article{Garcia2022INGRID,
  author  = {Garcia, B. M. and Umansky, M. V. and Watkins, J. and Guterl, J. and Izacard, O.},
  title   = {{INGRID}: An interactive grid generator for 2D edge plasma modeling},
  journal = {Computer Physics Communications},
  volume  = {275},
  pages   = {108316},
  year    = {2022},
  doi     = {10.1016/j.cpc.2022.108316}
}

@article{Hirshman1983,
  author  = {Hirshman, S. P. and Whitson, J. C.},
  title   = {Steepest-descent moment method for three-dimensional magnetohydrodynamic equilibria},
  journal = {The Physics of Fluids},
  volume  = {26},
  number  = {12},
  pages   = {3553--3568},
  year    = {1983},
  doi     = {10.1063/1.864116}
}

@article{Sauter2013,
  title     = {Tokamak coordinate conventions: {COCOS}},
  volume    = {184},
  issn      = {0010-4655},
  doi       = {10.1016/j.cpc.2012.09.010},
  number    = {2},
  journal   = {Computer Physics Communications},
  publisher = {Elsevier BV},
  author    = {Sauter, O. and Medvedev, S.Yu.},
  year      = {2013},
  pages     = {293--302}
}

@article{Lutjens1996,
  title     = {The {CHEASE} code for toroidal {MHD} equilibria},
  volume    = {97},
  issn      = {0010-4655},
  doi       = {10.1016/0010-4655(96)00046-x},
  number    = {3},
  journal   = {Computer Physics Communications},
  publisher = {Elsevier BV},
  author    = {L{\"u}tjens, H. and Bondeson, A. and Sauter, O.},
  year      = {1996},
  pages     = {219--260}
}

@article{Pataki2013,
  title     = {A fast, high-order solver for the Grad--Shafranov equation},
  volume    = {243},
  issn      = {0021-9991},
  doi       = {10.1016/j.jcp.2013.02.045},
  journal   = {Journal of Computational Physics},
  publisher = {Elsevier BV},
  author    = {Pataki, Andras and Cerfon, Antoine J. and Freidberg, Jeffrey P. and Greengard, Leslie and O'Neil, Michael},
  year      = {2013},
  pages     = {28--45}
}

@article{Lee2015,
  title     = {{ECOM}: A fast and accurate solver for toroidal axisymmetric {MHD} equilibria},
  volume    = {190},
  issn      = {0010-4655},
  doi       = {10.1016/j.cpc.2015.01.015},
  journal   = {Computer Physics Communications},
  publisher = {Elsevier BV},
  author    = {Lee, Jungpyo and Cerfon, Antoine},
  year      = {2015},
  pages     = {72--88}
}

@article{Lao1985,
  title     = {Reconstruction of current profile parameters and plasma shapes in tokamaks},
  volume    = {25},
  issn      = {1741-4326},
  doi       = {10.1088/0029-5515/25/11/007},
  number    = {11},
  journal   = {Nuclear Fusion},
  publisher = {IOP Publishing},
  author    = {Lao, L.L. and St. John, H. and Stambaugh, R.D. and Kellman, A.G. and Pfeiffer, W.},
  year      = {1985},
  pages     = {1611--1622}
}

@misc{Xie2026arxiv,
  author        = {Xie, Huasheng and Li, Yueyan},
  title         = {What Is the Minimum Number of Parameters Required to Represent Solutions of the {Grad--Shafranov} Equation?},
  year          = {2026},
  doi           = {10.48550/arXiv.2601.02942},
  note          = {arXiv:2601.02942}
}

@misc{Li2026VEQR,
  author        = {Li, Xingyu and Xie, Huasheng and Wei, Lai and Wang, Zhengxiong},
  title         = {Investigation of Toroidal Rotation Effects on Spherical Torus Equilibria using the Fast Spectral Solver {VEQ-R}},
  year          = {2026},
  doi           = {10.48550/arXiv.2602.11422},
  note          = {arXiv:2602.11422}
}

@article{Lao1981,
  title     = {Variational moment solutions to the Grad--Shafranov equation},
  volume    = {24},
  issn      = {0031-9171},
  doi       = {10.1063/1.863562},
  number    = {8},
  journal   = {The Physics of Fluids},
  publisher = {AIP Publishing},
  author    = {Lao, L. L. and Hirshman, S. P. and Wieland, R. M.},
  year      = {1981},
  pages     = {1431--1440}
}

@article{Lao1982,
  title     = {{VMOMS} -- A computer code for finding moment solutions to the Grad--Shafranov equation},
  volume    = {27},
  issn      = {0010-4655},
  doi       = {10.1016/0010-4655(82)90069-8},
  number    = {2},
  journal   = {Computer Physics Communications},
  publisher = {Elsevier BV},
  author    = {Lao, L.L. and Wieland, R.M. and Houlberg, W.A. and Hirshman, S.P.},
  year      = {1982},
  pages     = {129--146}
}

@article{Lao1984,
  title     = {Variational moment method for computing magnetohydrodynamic equilibria},
  volume    = {31},
  issn      = {0010-4655},
  doi       = {10.1016/0010-4655(84)90045-6},
  number    = {2-3},
  journal   = {Computer Physics Communications},
  publisher = {Elsevier BV},
  author    = {Lao, L.L.},
  year      = {1984},
  pages     = {201--212}
}

@article{Haney1995,
  title     = {A fast, user-friendly code for calculating magnetohydrodynamic equilibria},
  volume    = {9},
  issn      = {0894-1866},
  doi       = {10.1063/1.168526},
  number    = {2},
  journal   = {Computers in Physics},
  publisher = {AIP Publishing},
  author    = {Haney, S. W. and Freidberg, J. P. and Solomon, C. J.},
  year      = {1995},
  pages     = {216--224}
}

@article{Ludwig1995,
  title     = {Direct variational solutions to the Grad--Schluter--Shafranov equation},
  volume    = {37},
  issn      = {1361-6587},
  doi       = {10.1088/0741-3335/37/6/003},
  number    = {6},
  journal   = {Plasma Physics and Controlled Fusion},
  publisher = {IOP Publishing},
  author    = {Ludwig, G O},
  year      = {1995},
  pages     = {633--646}
}

@article{Miller1998,
  title     = {Noncircular, finite aspect ratio, local equilibrium model},
  volume    = {5},
  issn      = {1089-7674},
  doi       = {10.1063/1.872666},
  number    = {4},
  journal   = {Physics of Plasmas},
  publisher = {AIP Publishing},
  author    = {Miller, R. L. and Chu, M. S. and Greene, J. M. and Lin-Liu, Y. R. and Waltz, R. E.},
  year      = {1998},
  pages     = {973--978}
}

@article{Arbon2021,
  title     = {Rapidly-convergent flux-surface shape parameterization},
  volume    = {63},
  issn      = {1361-6587},
  doi       = {10.1088/1361-6587/abc63b},
  number    = {1},
  journal   = {Plasma Physics and Controlled Fusion},
  publisher = {IOP Publishing},
  author    = {Arbon, R and Candy, J and Belli, E A},
  year      = {2021},
  pages     = {012001}
}

@article{Snoep2023,
  title     = {Improved flux-surface parameterization through constrained nonlinear optimization},
  volume    = {30},
  issn      = {1089-7674},
  doi       = {10.1063/5.0145001},
  number    = {6},
  journal   = {Physics of Plasmas},
  publisher = {AIP Publishing},
  author    = {Snoep, G. and Koenders, J. T. W. and Bourdelle, C. and Citrin, J.},
  year      = {2023},
  pages     = {063906}
}

@article{Lewis1990,
  title     = {Physical constraints on the coefficients of Fourier expansions in cylindrical coordinates},
  volume    = {31},
  issn      = {1089-7658},
  doi       = {10.1063/1.529009},
  number    = {11},
  journal   = {Journal of Mathematical Physics},
  publisher = {AIP Publishing},
  author    = {Lewis, H. Ralph and Bellan, Paul M.},
  year      = {1990},
  pages     = {2592--2596}
}

@article{Cerfon2010,
  title     = {``One size fits all'' analytic solutions to the Grad--Shafranov equation},
  volume    = {17},
  issn      = {1089-7674},
  doi       = {10.1063/1.3328818},
  number    = {3},
  journal   = {Physics of Plasmas},
  publisher = {AIP Publishing},
  author    = {Cerfon, Antoine J. and Freidberg, Jeffrey P.},
  year      = {2010},
  pages     = {032502}
}

@book{Trefethen2000,
  title     = {Spectral Methods in MATLAB},
  isbn      = {9780898719598},
  doi       = {10.1137/1.9780898719598},
  publisher = {Society for Industrial and Applied Mathematics},
  author    = {Trefethen, Lloyd N.},
  year      = {2000}
}

@article{Artaud2018,
  author  = {Artaud, J. F. and others},
  title   = {{METIS}: a fast integrated tokamak modelling tool for scenario design},
  journal = {Nuclear Fusion},
  volume  = {58},
  number  = {10},
  pages   = {105001},
  year    = {2018},
  doi     = {10.1088/1741-4326/aad5b1},
}

@article{Imbeaux2015,
  author  = {Imbeaux, F. and Pinches, S. D. and Lister, J. B. and others},
  title   = {Design and first applications of the {ITER} Integrated Modelling \& Analysis Suite},
  journal = {Nuclear Fusion},
  volume  = {55},
  number  = {12},
  pages   = {123006},
  year    = {2015},
  doi     = {10.1088/0029-5515/55/12/123006},
}

@article{Felici2018,
  author  = {Felici, F. and Citrin, J. and Teplukhina, A. and Redondo, J. and Bourdelle, C. and Imbeaux, F. and Sauter, O. and others},
  title   = {Real-time-capable prediction of temperature and density profiles in a tokamak using {RAPTOR} and a first-principle-based transport model},
  journal = {Nuclear Fusion},
  volume  = {58},
  number  = {9},
  pages   = {096006},
  year    = {2018},
  doi     = {10.1088/1741-4326/aac8f0},
}

@article{Virtanen2020,
  author  = {Virtanen, Pauli and others},
  title   = {{SciPy} 1.0: fundamental algorithms for scientific computing in {Python}},
  journal = {Nature Methods},
  volume  = {17},
  number  = {3},
  pages   = {261--272},
  year    = {2020},
  doi     = {10.1038/s41592-019-0686-2},
}

@inproceedings{Lam2015,
  series     = {SC15},
  title      = {{Numba}: an {LLVM}-based {Python} {JIT} compiler},
  doi        = {10.1145/2833157.2833162},
  booktitle  = {Proceedings of the Second Workshop on the LLVM Compiler Infrastructure in HPC},
  publisher  = {ACM},
  author     = {Lam, Siu Kwan and Pitrou, Antoine and Seibert, Stanley},
  year       = {2015},
  pages      = {1--6},
  collection = {SC15}
}

@article{Haney1992,
  title     = {A ``SuperCode'' for Systems Analysis of Tokamak Experiments and Reactors},
  volume    = {21},
  issn      = {0748-1896},
  doi       = {10.13182/fst92-a29974},
  number    = {3P2A},
  journal   = {Fusion Technology},
  publisher = {Informa UK Limited},
  author    = {Haney, S. W. and Barr, W. L. and Crotinger, J. A. and Perkins, L. J. and Solomon, C. J. and Chaniotakis, E. A. and Freidberg, J. P. and Wei, J. and Galambos, J. D. and Mandrekas, J.},
  year      = {1992},
  pages     = {1749--1758}
}

@article{Dudt2020,
  title     = {{DESC}: A stellarator equilibrium solver},
  volume    = {27},
  issn      = {1089-7674},
  doi       = {10.1063/5.0020743},
  number    = {10},
  journal   = {Physics of Plasmas},
  publisher = {AIP Publishing},
  author    = {Dudt, D. W. and Kolemen, E.},
  year      = {2020},
  pages     = {102513}
}

@article{Amorisco2024,
  title     = {{FreeGSNKE}: A {Python}-based dynamic free-boundary toroidal plasma equilibrium solver},
  volume    = {31},
  issn      = {1089-7674},
  doi       = {10.1063/5.0188467},
  number    = {4},
  journal   = {Physics of Plasmas},
  publisher = {AIP Publishing},
  author    = {Amorisco, N. C. and Agnello, A. and Holt, G. and Mars, M. and Buchanan, J. and Pamela, S. J. P.},
  year      = {2024},
  pages     = {042517}
}

@inproceedings{Kripner2019PLEQUE,
  author    = {Kripner, L. and Tome{\v{s}}, M. and Peterka, M. and Urban, J. and Mac{\'u}{\v{s}}ov{\'a}, E. and Jaulmes, F. and Fridrich, D. and Grover, O. and Ficker, O. and Krbec, J. and {\v{C}}erovsk{\'y}, J.},
  title     = {Towards the Integrated Analysis of Tokamak Plasma Equilibria: {PLEQUE}},
  booktitle = {{WDS'19 Proceedings of Contributed Papers --- Physics}},
  pages     = {100--105},
  year      = {2019},
  isbn      = {978-80-7378-409-6},
}

@book{Nocedal2006,
  author    = {Nocedal, Jorge and Wright, Stephen J.},
  title     = {Numerical Optimization},
  edition   = {2},
  publisher = {Springer},
  address   = {New York},
  series    = {Springer Series in Operations Research and Financial Engineering},
  year      = {2006},
  isbn      = {978-0-387-30303-1},
  doi       = {10.1007/978-0-387-40065-5}
}

@misc{FreeQDSKGeqdsk,
  author       = {{FreeQDSK developers}},
  title        = {{GEQDSK} file format documentation},
  year         = {2026},
  howpublished = {{FreeQDSK} 0.5.2 documentation},
  note         = {Accessed 2026-05-07}
}

\end{document}